\documentclass[reqno, 11pt]{amsart}

\usepackage{hyperref}

\usepackage{enumerate}
\newcommand\bx{{\mathbf x}}
\newcommand\by{{\mathbf y}}
\newcommand\bz{{\mathbf z}}
\newcommand\bq{{\mathbf q}}
\newcommand\bp{{\mathbf p}}
\newcommand\be{{\mathbf e}}

\newcommand\bk{{\mathbf k}}
\newcommand\bu{{\mathbf u}}
\newcommand\bv{{\mathbf v}}

\newcommand\E{{\mathbb E}}

\newcommand\R{{\mathbb R}}
\newcommand\To{{\mathbb Z}}
\newcommand\Z{{\mathbb Z}}
\newcommand\cE{\mathcal E}

\newcommand\CE{{\mathbb E}_{\scriptscriptstyle{N,T}}}
\newcommand\ME{{\mathbb E}_{\scriptscriptstyle{N,\mathcal E}}}

\usepackage{amssymb, amsmath}

\newtheorem{theo}{Theorem}
\newtheorem{lemma}{Lemma}

\title[Thermal conductivity]{%
Thermal conductivity for a momentum conservative model}
\author{Giada Basile}
\address{\!\!\!\!\!\!\!\! WIAS,\newline
Mohrenstr.39, \newline
10117 Berlin - Germany.\newline
\rm {\texttt{basile@wias-berlin.de}} 
}

\author{C\'edric Bernardin}
\address{\!\!\!\!\!\!\!Universit\'e de Lyon, CNRS (UMPA)\newline
Ecole Normale Sup\'erieure de Lyon,\newline
46, all\'ee d'Italie,\newline
 69364 Lyon Cedex 07 - France.\newline
\rm {\texttt{Cedric.Bernardin@umpa.ens-lyon.fr}}\newline
\texttt{http://w3umpa.ens-lyon.fr/\~\;\!\!cbernard/}
}

\author{Stefano Olla}
\address{\!\!\!\!\!\!\!Ceremade, UMR CNRS 7534 \newline
  Universit\'e de Paris Dauphine,\newline
  Place du Mar\'echal De Lattre De Tassigny\newline
  75775 Paris Cedex 16 - France.\newline
\rm {\texttt{olla@ceremade.dauphine.fr}}\newline
 \texttt{http://www.ceremade.dauphine.fr/\~\;\!\!olla}
}

\date{\today}

\thanks{\textsc{Acknowledgements.}
We acknowledge the support of the
 ANR LHMSHE n.BLAN07-2184264 and of the accord GREFI-MEFI}

%Use showkeys class
\begin{document}

%\preprint{APS/123-QED}
\begin{abstract}
% We present here  complete mathematical proofs of the results annouced
% in cond-mat/0509688.
% Anomalous large thermal conductivity has been observed numerically and
% experimentally in one and two dimensional systems. All explicitly
% solvable microscopic models proposed until now did not explain this
% phenomenon and there is an open debate about the role of conservation
% of momentum.
We introduce a model whose thermal conductivity diverges 
 in dimension 1 and 2, while it remains finite in dimension 3. 
We consider a system of  oscillators perturbed by a stochastic
dynamics conserving  momentum and energy. We compute
thermal conductivity via Green-Kubo formula.
In the harmonic case
we compute the current-current time correlation function, that
 decay like
$t^{-d/2}$  in the unpinned case and like $t^{-d/2-1}$ if a on-site
harmonic potential is present. This implies a finite conductivity  in
$d\ge 3$ or in pinned cases, and we compute it explicitely.
For general anharmonic strictly convex interactions we prove some upper
bounds  for the conductivity that behave qualitatively as in the
harmonic cases.
\end{abstract}

\maketitle

\section{Introduction}
\label{sec:intro}

The mathematical deduction of Fourier's law and heat equation for the
diffusion of energy from a microscopic Hamiltonian deterministic
dynamics is one of the major open problem in non-equilibrium
statistical mechanics \cite{blr}. Even the existence of the thermal
conductivity defined by Green-Kubo formula, is a challenging
mathematical problem and it may be infinite in some low dimensional
cases \cite{sll}. Let us consider the problem in a generic lattice
system where dynamics conserves energy (between other quantities like
momentum etc.). For $\bx \in \mathbb Z^d$, denote by ${\cE}_{\bx}(t)$ the
energy of \emph{atom} $\bx$. To simplify notations let us consider the
1-dimensional case. Since the dynamics conserves the total
energy, there exist \emph{energy currents} $j_{x, x+1}$
(local functions of the coordinates of the system), such that
\begin{equation}\label{in:01}
  \frac d{dt} \cE_{x}(t) = j_{x-1, x}(t) - j_{x, x+1}(t) 
\end{equation}
Another consequence of the conservation of energy is that there %will
exists a family of stationary equilibrium measures parametrized by
 temperature value $T$ (between other possible parameters). 
Let us denote by $<\cdot> = <\cdot>_T$ the expectation of the system
 starting from this equilibrium measure, and assume that parameters
 are set so that $<j_{x,y}>=0$ (for example if total momentum
 is fixed to be null).   Typically these measures are
 Gibbs measure with sufficiently fast decay of space correlations so that
 energy has static fluctuation that are Gaussian distributed if
 properly rescaled in space. 
%Defining the space-time correlations of
% the energy as
% \begin{equation*}
%   S(x,t) = <\cE_x(t) {\cE}_{0}(0)> - <{\cE}_{0}>^2 
% \end{equation*}
% then static fluctuations of energy, under the equilibrium measure,
% have variance given by
% \begin{equation*}
%   \chi = \sum_{x \in {\mathbb Z}} x^2 S(x,0)
% \end{equation*}
%If thermal conductivity is finite, the energy fluctuations should
%evolve in time (in a proper space-time large scale) as a diffusion,  
Let us define the  space-time correlations of the energy as
\begin{equation*}
S(x,t) = <\cE_x(t) {\cE}_{0}(0)> - <{\cE}_{0}>^2 .
 \end{equation*}
If thermal conductivity is finite, $S(x,t)$ should be solution of the
diffusion equation
(in a proper large space-time scale) 
and  thermal conductivity (TC) can be defined as 
\begin{equation}\label{eq:variance}
  \kappa(T) = \lim_{t\to\infty} \frac{1}{2t T^2} \sum_{x \in \mathbb Z} x^2 S(x,t).
\end{equation}
By using the energy conservation law (\ref{in:01}), time and space
invariance (see section \ref{sec:gkf}), one can rewrite
\begin{equation}\label{in:gk}
  \begin{split}
    \kappa(T) &= \lim_{t\to\infty} \frac{1}{2t T^2} \sum_{x \in {\mathbb Z}} \left<
      \left(\int_0^t j_{x,x+1}(s) ds\right) \left(\int_0^t j_{0,1}(s')
        ds'\right) \right>\\
    &= \frac{1}{ T^2} \sum_{x \in {\mathbb Z}} \int_0^\infty \left<
      j_{x,x+1}(t)\, j_{0,1}(0)\right>\,
        dt
  \end{split}
\end{equation}
which is the celebrated Green-Kubo formula for the thermal
conductivity (cf. \cite{spohn}). 

One can see from (\ref{in:gk}) why the problem is so difficult for
deterministic dynamics: one needs some control of time decay of
the current-current correlations, a difficult problem even for finite
dimensional dynamical systems.  
Furthermore in some one--dimensional systems, like Fermi-Pasta-Ulam
chain of unpinned oscillators, if total momentum is
conserved by the dynamics, thermal conductivity is expected to be
infinite (cf. \cite{sll} for a review of numerical results on this topic).
Very few mathematically rigorous results exist for deterministic
systems (\cite{rll, bk}).

In this paper we consider stochastic perturbations of a deterministic
Hamiltonian dynamics on a multidimensional lattice and we study the
corresponding thermal conductivity as defined by (\ref{in:gk}). The stochastic
perturbations are such that they exchange momentum between 
particles with a local random mechanism that conserves total energy and total
momentum. 

Thermal conductivity of Hamiltonian systems with stochastic dynamical
perturbations have been studied for harmonic chains. In  \cite{brv,
  bll} the stochastic perturbation does not conserve energy, and in
\cite{bo} only energy is conserved. The novelty of our work is that
our stochastic perturbations conserve also momentum, with dramatic
consequences in low dimensional systems. In fact we prove that for
unpinned systems (where also the Hamiltonian dynamics conserve
momentum, see next section for a precise definition) with
harmonic interactions, thermal conductivity is infinite in 1 and 2
dimensions, while is finite for $d\ge 3$ or for pinned systems. 
Notice that for stochastic
perturbations of harmonic systems that do not conserve momentum,
thermal conductivity is always finite \cite{bll, bo}.  

This divergence of TC in dimension 1 and 2 is
expected generically for deterministic Hamiltonian non-linear system
when unpinned.
%(cf. \cite{sll} for a general review of the physical
%literature). 
So TC in our model behaves qualitatively
like in a deterministic non-linear system, i.e. these stochastic
interactions reproduce some of the features of the non-linear
deterministic hamiltonian interactions. 
Also notice that because of the conservation
laws, the noise that we introduce is of multiplicative type,
i.e. intrinsically non-linear (cf. (\ref{eq:sde}) and
(\ref{eq:sde1})). 
On the other hand, purely deterministic harmonic chains (pinned or
unpinned and in any dimension) have always
infinite conductivity \cite{rll}. In fact in these linear systems energy
fluctuations are transported ballistically by waves that do not
interact with each other. Consequently, in the harmonic case,
 our noise is entirely
responsable for the finiteness of the TC in dimension 3 and for pinned
systems. Also in dimension 1 and 2, the divergence of TC for  
unpinned harmonic
systems is due to a superdiffusion of the energy fluctuations, not to
ballistic transport (see \cite{bos, jko} where this behavior is explained
with a kinetic argument).

For anharmonic systems, even with the stochastic noise we are not able
to prove the existence of thermal conductivity (finite or
infinite). If the dimension $d$ is greater than $3$ and the system is
pinned, we get a uniform bound on the finite size system
conductivity. For low dimensional pinned systems ($d=1,2$), we can
show the conductivity is finite if the interaction potential is
quadratic and the pinning is generic. For the unpinned system we have
to assume that the interaction between nearest-neighbor particles is
strictly convex and quadratically bounded at infinity. This because we
need some informations on the spatial decay of correlations in the
stationary equilibrium measure, that decay slow in unpinned system
\cite{dd}. In this case, we prove the conductivity is finite in
dimension $d\geq 3$ and we obtain upper bounds in the size $N$ of the
system of the form $\sqrt{N}$ in $d=1$ and $(\log N)^2$ in $d=2$
(see Theorem \ref{th-anharm} for precise statements).  

The paper is organized as follows. Section \ref{sec:dynamics} is
devoted to the precise description of the dynamics. In section
\ref{sec:gkf}, we present our results. The proofs of the harmonic case are
in section \ref{sec:cfe} and \ref{harm2} while the proofs of the
anharmonic case are stated in section
\ref{sec:anharm-case:-bounds}. The final section contains technical
lemmas related to equivalence of ensembles.

\vspace{0,5cm}
\textbf{Notations} : The canonical basis of $\R^d$ is noted
$(\be_1,\be_2, \ldots,\be_d)$ and the coordinates of a vector $\bu \in
\R^d$ are noted $(\bu^1,\ldots,\bu^d)$. Its Euclidian norm $|\bu|$  is
given by $|\bu|=\sqrt{(\bu^1)^2+ \ldots +(\bu^d)^2}$ and the scalar
product of $\bu$ and $\bv$ is $\bu \cdot \bv$.  

If $N$ is a positive integer, $\mathbb Z_N^d$ denotes the $d$-dimensional 
discrete torus of length $N$ and we identify $\bx = \bx + kN\be_j$
for any $j= 1,\dots,d$ and $k\in \mathbb Z$.  

If $F$ is a function from ${\mathbb Z}^d$ (or $\To_N^d$) into $\mathbb
R$ then the (discrete) gradient of $F$ in the direction $\be_j$ is
defined by $(\nabla_{\be_j} F)(\bx)= F(\bx +\be_j) -F(\bx)$ and the
Laplacian of $F$ is given by $(\Delta F) (\bx)= \sum_{j=1}^d
\left\{F(\bx +\be_j)+F(\bx-\be_j)-2F(\bx)\right\}$. 

% The microcanonical measure on $\To_N^d$ with mean energy $e$ is
% denoted by $<\cdot>_{\scriptscriptstyle{N,e}}$ and the canonical
% measure with temperature $T$ by
% $<\cdot>_{\scriptscriptstyle{N,T}}$. The subscript $e$ (resp. $T$)
% will be exclusively used in the microcanonical (resp. canonical)
% setting.  

\section{The dynamics}
\label{sec:dynamics}

In order to avoid difficulties with definitions of the dynamics and
its stationary Gibbs measures, we start with a finite system and we
will define thermal conductivity through an infinite volume limit
procedure (see section \ref{sec:gkf}).

We consider  the dynamics of the system of length $N$ with 
periodic boundary conditions. The atoms are labeled by $\bx \in
\mathbb Z_N^d$. Momentum of atom $\bx$ is $\bp_\bx \in {\R^d}$ and its displacement from its equilibrium position is ${\bq_x} \in \R^d$. 
The Hamiltonian is given by
\begin{equation}
  \label{eq:hamilt}
  \mathcal H_N =  \sum_{\bx \in \To_N^d}  \left[ \frac{|\bp_\bx|^2}2 + W(\bq_\bx)
    + \cfrac{1}{2} \sum_{|\by-\bx|=1} V(\bq_\bx - \bq_\by) \right] . \nonumber 
\end{equation}
% . We denote with $\nabla$, $\nabla^*$ and 
% $\Delta = \nabla^*\cdot \nabla$
% respectively the discrete gradient, its adjoint and the discrete Laplacian  on
% $\mathbb Z_N^d$. These are defined as
% \begin{equation}
%   \label{eq:grad}
%   \nabla_{\be_j} f(\bx) = f(\bx + \be_j) - f(\bx)
% \end{equation}
% and
% \begin{equation}
%   \label{eq:adjgrad}
%    \nabla^*_{\be_j} f(\bx) = f(\bx) - f(\bx - \be_j) 
% \end{equation}
% $\{\bq_\bx\}$ are the displacements of the atoms 
% from their equilibrium positions.
% % The canonical basis of ${\mathbb R}^d$ is denoted by
% % ${\be}_1,\ldots,{\be}_d$.
% % The configuration of the system is then given by the positions and the
% % momenta of the particles $\{\bq_\bx, \bp_\bx\}$, where  ${\bf
% % q}_\bx=(q_\bx^1,\ldots,q_\bx^d) \in {\mathbb R}^d$ and ${\bf
% % p}_\bx=(p_\bx^1,\ldots,p_\bx^d) \in {\mathbb R}^d$. The position
% % $\bq_\bx$ should be interpreted here as the displacement of the atom
% % $\bx$ from an equilibrium position (that can be $\bx$ itself).
%  The parameter $\alpha > 0$ is the strength of the
% interparticles springs, and $\nu \ge 0$ is the strength of the pinning
% (on-site potential).

We assume  that $V$ and $W$ have the following form: 
$$ 
V(\bq_\bx - \bq_\by) = \sum_{j=1}^d V_j( q_\bx^j - q_\by^j), \qquad
W(\bq_\bx) = \sum_{j=1}^d W_j(q_\bx^j).
$$
and that $V_j, W_j$ are smooth and even. We call $V$ the interaction potential, and $W$ the pinning
potential. The case where $W=0$ will be called unpinned. 

We consider the stochastic dynamics generated by the operator
\begin{equation}
  \label{eq:10}
  L \; = \; A + \gamma S \; .
\end{equation}
The operator $A$ is the usual Hamiltonian vector field
 \begin{equation*}
  \label{eq:Agen}
  \begin{split}
    A =& \sum_\bx \left\{ \bp_\bx \cdot \partial_{\bq_\bx} -
     \partial_{\bq_\bx}\mathcal H_N \cdot  \partial_{\bp_\bx} \right\}
  % =& \sum_\bx \sum_{j=1}^d  \left\{ p^j_\bx \partial_{q^j_\bx} +
%     (\alpha \Delta - \nu I)
%     q^j_\bx  \partial_{p^j_\bx} \right\}
\end{split}
\end{equation*}
while $S$ is the generator of the stochastic perturbation and 
$\gamma > 0$ is a positive parameter that regulates its strength.
The operator $S$ acts only on the momentums $\{\bp_\bx\}$ and
generates a diffusion on the surface of constant kinetic energy and
constant momentum. This is defined as follows. If $d\ge2$, for every nearest
neighbor atoms $\bx$ and $\bz$, consider the $d-1$ dimensional surface 
of constant kinetic energy and momentum
\begin{equation*}
  \mathbb S_{e,\bp} \; = \; \left\{(\bp_\bx,\bp_\bz)\in \mathbb R^{2d}:
    \frac 12\left(|\bp_\bx|^2 + |\bp_\bz|^2\right) = e\; ; \; 
    \bp_\bx + \bp_\bz = \bp  \right\}\ .
\end{equation*}
The following vector fields are tangent to $ \mathbb S_{e,\bp}$
\begin{equation*}
  \label{eq:Xfield}
   X^{i,j}_{\bx, \bz} = (p^j_\bz-p^j_\bx) (\partial_{p^i_\bz}
   - \partial_{p^i_\bx})  -(p^i_\bz-p^i_\bx) (\partial_{p^j_\bz}
   - \partial_{p^j_\bx}) .
\end{equation*}
so $ \sum_{i,j =1}^d (X^{i,j}_{\bx, \bz})^2$ generates a diffusion on
$ \mathbb S_{e,\bp}$ (see \cite{iw}). 
In  $d\ge 2$ we define
\begin{equation*}
  \label{eq:Sgen}\begin{array}{ll}
  S & = \displaystyle\frac 1{2(d-1)} \sum_{\bx}\sum_{i,j,k}^d
  \left( X^{i,j}_{\bx, \bx+\be_k}\right)^2 \\
  & = \displaystyle\frac  1{4(d-1)} \sum_{\bx, \bz \in \mathbb Z^d_N \atop |\bx -
 \bz|=1}\sum_{i,j}
   \left( X^{i,j}_{\bx, \bz}\right)^2
\end{array}
\end{equation*}
where  ${\be}_1,\ldots,{\be}_d$ is canonical basis of ${\mathbb Z}^d$.

Observe that this noise conserves the total momentum $\sum_\bx
\bp_\bx$ and energy $\mathcal H_N$, i.e.  
\begin{equation*}
  S \; \sum_\bx \bp_\bx = 0\ ,\quad S\; \mathcal H_N = 0
\end{equation*}

In dimension 1, in order to conserve total momentum and total kinetic energy, 
we have to consider a random exchange of momentum between three 
consecutive atoms (because if $d=1$, ${\mathbb S}_{e,\bp}$ has
dimension $0$), and we define
$$
S = \frac 16 \sum_{x\in\mathbb{T}^1_N}(Y_x)^2
$$
where
\begin{equation*}
\label{eq:005}
Y_x=(p_x-p_{x+1})\partial_{p_{x-1}}+(p_{x+1}-p_{x-1})\partial_{p_x} +
(p_{x-1}-p_x)\partial_{p_{x+1}} 
\end{equation*}
which is vector field tangent to the surface of constant energy and
momentum of the three particles involved.

The corresponding Fokker-Planck equation for the time evolution of the
probability distribution $P(\bq, \bp, t)$, given an initial
distribution  $P(\bq, \bp, 0)$ is given by
\begin{equation}\label{FP}
 \frac {\partial P}{\partial t} = (- A + \gamma S) P =  L^* P\ .
\end{equation}
where $L^*$ is the the adjoint of $L$ with respect to the Lebesgue measure.

Let  $\{w^{i,j}_{\bx, \by} ;\; \bx, \by \in \mathbb
T_N^d;\; i,j= 1,\dots, d;\; |\by - \bx| = 1\}$ be independent standard
Wiener processes, such that $w^{i,j}_{\bx, \by} = w^{i,j}_{\by, \bx}$. 
Equation (\ref{FP}) corresponds to the law at time
$t$ of the solution of the following stochastic differential
equations: 
\begin{equation}
  \label{eq:sde}
  \begin{split}
     d\bq_\bx &= \bp_\bx\; dt\\
     d\bp_\bx &= - \partial_{\bq_\bx} \mathcal H_N\; dt +
     2 \gamma  \Delta \bp_\bx \; dt \\
     & \qquad \qquad \qquad + \frac {\sqrt{\gamma}}{2\sqrt{ d-1}}
      \sum_{\bz: |\bz - \bx|=1} \sum_{i,j=1}^d \left(X^{i,j}_{\bx, \bz}
     \bp_{\bx}  \right) \; dw^{i,j}_{\bx, \bz}(t) 
  \end{split}
\end{equation}
In $d=1$ these are:
\begin{equation}
  \label{eq:sde1}
  \begin{split}
    dp_x = -\partial_{q_x} H_N\; dt + \frac \gamma 6
    \Delta(4p_x + p_{x-1} + p_{x+1}) dt \\
    + \sqrt{\frac \gamma 3}
    \sum_{k=-1,0, 1} \left(Y_{x+k} p_{x} \right) dw_{x+k}(t)
  \end{split}
\end{equation}
where here $\{w_{x}(t), x= 1, \dots, N\}$ are independent standard
Wiener processes.

Defining the energy of the atom $\bx$ as
\begin{equation*}
  \label{eq:energyx}
  {\cE}_{\bx} = \frac 12  \bp_\bx^2 \, +\,  {W( \bq_\bx)} \,+\,
\cfrac{1}{2}\sum_{\by: |\by -\bx|=1}
V(\bq_{\by} - \bq_\bx) 
\end{equation*}
the energy conservation law can be read locally as
\begin{equation*}
   {\cE}_{\bx}(t) -   {\cE}_{\bx}(0) =  \sum_{k=1}^d \left(\,
J_{{\bx-\be_k, \bx}}([0,t]) - J_{{\bx,\bx +\be_k}}([0,t])\,\right)
 \end{equation*}
where $J_{\bx,\bx +\be_k}([0,t])$ is the total energy current
between $\bx$ and $\bx +\be_k$ up to
time $t$. This can be written as
\begin{equation}
  \label{eq:tc}
  J_{\bx, \bx +\be_k}({[0,t]})=\int_0^t j_{\bx, \bx +\be_k}(s) \; ds +
  M_{\bx, \bx +\be_k}(t) 
\end{equation}
In the above $M_{\bx, \bx +\be_k}(t)$ are  martingales that can be
written explicitly as It\^o stochastic integrals
\begin{equation}
  \label{eq:mart}
  M_{\bx, \bx +\be_k}(t) = \sqrt{\frac{\gamma}{(d-1)}} \sum_{i,j}
  \int_0^t \left(X^{i,j}_{\bx, \bx +\be_k} {\cE}_{\bx}\right)(s) \;
  dw^{i,j}_{\bx, \bx +\be_k} (s) 
\end{equation}
% where $\{w^{i,j}_{\bx, \bx +\be_k};\; \bx\in \mathbb Z_N^d;\;  i,j,k = 1,
% \dots, d\}$ are independent standard Wiener processes.

In $d=1$ these martingales write explicitly as
\begin{equation}
  \label{eq:mart1d}
  M_{x, x+1} (t)=  \sqrt{\frac{\gamma}{3}}  \int_0^t 
   \sum_{k=-1,0, 1} \left(Y_{x+k} \cE_{x} \right) dw_{x+k}(t)
\end{equation}

The instantaneous energy
currents $j_{\bx,\bx +\be_k}$ satisfy the equation 
\begin{equation*}
  L {\cE}_{\bx} = \sum_{k=1}^d \left(
j_{\bx-\be_k, \bx} - j_{\bx,\bx +\be_k}\right)
\end{equation*}
and it can be written as
\begin{equation}
  \label{eq:1}
  j_{\bx, \bx +\be_k} = j^{a}_{\bx, \bx +
\be_k} +\gamma j_{\bx, \bx +\be_k}^s \quad .
\end{equation}
The first term in (\ref{eq:1}) is the Hamiltonian contribution to the
energy current
\begin{equation}
  \label{eq:2}
  \begin{split}
    j^a_{\bx,\bx +\be_k} &= -\cfrac{1}{2} (\nabla V)(\bq_{\bx+\be_k} -
    \bq_\bx)\cdot (\bp_{\bx+\be_k} + \bp_\bx)\\
    &= -\frac 12 \sum_{j=1}^d V'_j(q^j_{\bx+\be_k} -
    q^j_\bx) (p^j_{\bx+\be_k} + p^j_\bx)
  \end{split}
\end{equation}
while the noise contribution in $d\ge 2$ is
\begin{equation}
  \label{eq:3}
     \gamma j^s_{\bx,\bx +\be_k} =- \gamma(\nabla_{\be_k} \bp^2)_\bx
 \end{equation}
and in $d=1$ is
\begin{equation*}
\begin{split}
  \gamma j^s_{x, x + 1} =& -\gamma\nabla \varphi (p_{x-1},p_x,p_{x+1})
  \\
 \varphi (p_{x-1},p_x,p_{x+1})&=
\frac 16 [p_{x+1}^2 + 4 p_{x}^2 +
    p_{x-1}^2 +  p_{x+1} p_{x-1} -2 p_{x+1} p_{x}-2 p_{x} p_{x-1}]
\end{split}
\end{equation*}

In the unpinned case ($W=0$),
 given any values of $\mathcal E>0$,
the uniform probability measure on the constant energy-momentum
 shell
\begin{equation*}
 \Sigma_{N, \mathcal E} =  \left\{(\bp,\bq): \mathcal H_N = N \mathcal
   E, \; \sum_{\bx\in
     \mathbb Z_N^d} \bp_x = 0,  \sum_{\bx\in
     \mathbb Z_N^d} \bq_\bx = 0\right\} 
\end{equation*}
is stationary for the dynamics, and $A$ and $S$ are respectively
antisymmetric and symmetric with respect to this measure. 
For the stochastic dynamics, we believe
that these measures are also ergodic, 
i.e. total energy, total momentum and center of mass are the only
conserved quantities. Notice that, because of the periodic boundary
conditions, no other conserved quantities associated to
the distortion of the lattice exist. 
For example in $d=1$ the total length
of the chain $\sum_\bx (q_{x + 1} - q_x)$ is automatically null.

In the pinned case, total momentum is not conserved, and the ergodic
stationary measures are given by the uniform probability measures on
the energy shells
\begin{equation*}
 \Sigma_{N,\mathcal E} =  \left\{(\bp,\bq): \mathcal H_N = N\mathcal E
   \,\right\} . 
\end{equation*}

In both cases we refer to these measures as \emph{microcanonical} Gibbs
measures. We denote by $<\cdot>_{\scriptscriptstyle {N,\mathcal E}}$ the
expectation with respect to these microcanonical measures. 

We will also consider the dynamics starting from the canonical Gibbs
measure $<\cdot>_{\scriptscriptstyle {N,T}}$ with temperature $T>0$
defined on the phase space $(\mathbb R^{2d})^{\mathbb Z^d_N}$ by 
$$
<\cdot>_{\scriptscriptstyle {N,T}} = \frac {e^{-\mathcal H_N/T
  }}{Z_{\scriptscriptstyle{N,T}}} d\bq\, d \bp .
$$  
To avoid confusions between these measures we restrict the use of
subscript $\mathcal E$ for the microcanonical measure and subscript $T$ for the
canonical measure.

%%%%%%%%%%%%%%%%%%%%%%%%%%%%%%%%%%%%%%%%%%%%%%%%%%%%%%%%%%%%%%%%%%%%%%%
%%%%%%%%%%%%%%%%%%%%%%%%%%%%%%%%%%%%%%%%%%%%%%%%%%%%%%%%%%%%%%%%%%%%%%%
%%%%%%%%%%%%%%%%%%%%     GREEN-KUBO    %%%%%%%%%%%%%%%%%%%%%%%%%%%%%%%%
%%%%%%%%%%%%%%%%%%%%                   %%%%%%%%%%%%%%%%%%%%%%%%%%%%%%%%
%%%%%%%%%%%%%%%%%%%%%%%%%%%%%%%%%%%%%%%%%%%%%%%%%%%%%%%%%%%%%%%%%%%%%%%
%%%%%%%%%%%%%%%%%%%%%%%%%%%%%%%%%%%%%%%%%%%%%%%%%%%%%%%%%%%%%%%%%%%%%%%

\section{Green-Kubo formula and statement of the results}
\label{sec:gkf}
In the physical literature several variations of the Green-Kubo
formula \eqref{in:gk} can be found (\cite{sll, bll}). 
As in (\ref{in:gk}), one can start with the infinite
system and sum over all $x \in {\mathbb Z}^d$. 
One can also start working with the finite system with periodic boundary
conditions and sum over $x \in {\Lambda}_N^d$ where $\Lambda_N^d$ is a finite
box of size $N$ and take the thermodynamic limit $N \to \infty$
(before sending the time to infinity). 
In the finite case there is a choice of the equilibrium measure. If
$<\cdot>$ is the canonical measure at temperature $T$, 
one refers to the derivation
{\textit {\`a la}} Kubo.  If $<\cdot>$ is the microcanonical
measure at energy $\mathcal E N^d$,
 one refers to the derivation {\textit {\`a la}} Green. Because of
to the equivalence of ensembles one expects that these different
definitions give all the same value of the conductivity,  
provided that temperature 
$T$ and energy $\cE$ are suitably related by the
corresponding thermodynamis relation. Nevertheless
a rigorous justification is absent in the literature.  

In the sequel we will consider the microcanonical Green-Kubo formula
(noted $\kappa$) and the canonical Green-Kubo formula
(noted $\tilde \kappa$) starting from our finite system.

In the harmonic case we work out the microcanonical Green-Kubo
version that we compute explicitly. Similar computations are valid
(with less work) for the canonical version of the Green-Kubo
formula and will give the same result.
In the anharmonic case equivalence of ensembles is less developed and
we deal only with the canonical version of the Green-Kubo formula. \\

 % the uniform measure on the constant
% energy-momentum shell
% \begin{equation*}
%  \Sigma_{N,e} =  \left\{(\bp,\bq): \mathcal H_N = Ne, \; \sum_{\bx\in \mathbb Z_N^d}
%     \bp_x = 0 \right\}
% \end{equation*}
The microcanonical Green-Kubo formula for the conductivity in the
direction ${\be_1}$ is defined as the limit (when it exists) 
\begin{equation}
\label{eq:gk0}
%\begin{split}
\kappa^{1,1}(T)  =   \lim_{t\to\infty} \lim_{N\to\infty} \frac 1{2
  T^2 t} 
\sum_{\bx\in  \mathbb Z_N^d}
\ME \left[J_{\bx,\bx+\be_1}([0,t]) \, J_{0,\be_1}([0,t]) \right]%\\ 
%& = \lim_{t\to\infty}  \lim_{N\to\infty} \frac 1{2 e^2 t} \frac 1{N^{d}}\mathbb E
%\left(\left[\sum_{\bx\in \mathbb Z_N^d} J_{\bx,\bx+\be_1}([0,t])\right]^2 \right) 
%\end{split}
\end{equation}
where $\ME$ is the expectation starting with the microcanonical
distribution $<\cdot>_{\scriptscriptstyle{N,\mathcal E}}$,
and the energy  $\mathcal E = \mathcal E(T)$ is chosen such that it
corresponds to the  
thermodynamic energy at temperature $T$ (i.e. the average of the 
kinetic energy in the canonical measure).   In the harmonic case 
$T= \mathcal E$.

Similarly the canonical version of the Green-Kubo formula is given by
\begin{equation}
  \label{eq:gcc}
  \tilde \kappa^{1,1} (T) = \lim_{t\to \infty} \lim_{N\to \infty} 
   \frac 1{2 T^2 t} \sum_{\bx \in {\mathbb Z}_N^d} \CE 
   \left[ \,J_{\bx,\bx+\be_1}([0,t]) 
     J_{0,\be_1}([0,t]) \, \right] 
\end{equation}
when this limit exists. Here $\CE$ indicates the expectation with respect to the
equilibrium dynamics starting with the canonical measure 
$<\cdot>_{\scriptscriptstyle{N,T}}$ at temperature $T$. 
These definitions are consistent with
(\ref{eq:variance})-(\ref{in:gk}) as we show at the end of this
section. 

Our first results concern the \textit{$(\alpha,\nu)$-harmonic case}:
\begin{equation}
\label{eq:harm}
V_j (r)=\alpha r^2, \quad W_j (q)=\nu q^2, \quad \alpha >0, \quad \nu \geq 0
\end{equation}

\begin{theo}\label{th-harm}
In the $(\alpha,\nu)$-harmonic case (\ref{eq:harm}), the limits
defining $\kappa^{1,1}$ and ${\tilde \kappa}^{1,1}$ exist. They are
finite if $d\geq 3$ or if the on-site harmonic potential is present
($\nu>0$), 
and are infinite in the other cases. 
When finite, $\kappa(T)$ and ${\tilde \kappa} (T)$ are
independent of $T$, coincide and
the following formula holds  
\begin{equation}\label{cond1}
{\tilde \kappa }^{1,1} (T)=\kappa^{1,1}(T)= \cfrac{1}{8\pi^2 d \gamma}
\int_{[0,1]^d}\cfrac{(\partial_{{\bk}^1}\omega)^2(\bk)}{\psi
  (\bk)}d\bk + \cfrac{\gamma}{d}
\end{equation}
where $\omega(\bk)$ is the dispertion relation
\begin{equation}
\label{eq:dr}
\omega (\bk) = 
\left(\nu + 4\alpha\sum_{j=1}^{d} \sin^{2} (\pi \bk^j)\right)^{1/2}
\end{equation}
and
\begin{equation}
\label{eq:psi}
\psi (\bk) = 
\begin{cases}
8\sum_{j=1}^{d} \sin^{2} (\pi \bk^j), \quad \mbox{ if } \quad d\geq 2\\
4/3 \ \sin^2 (\pi \bk) (1+2\cos^2 (\pi \bk)), \quad \mbox{ if } \quad d=1
\end{cases}
\end{equation}
% in (\ref{eq:dr}) and (\ref{eq:psi}). 
\end{theo}

Consequently in the unpinned harmonic cases in dimension $d=1$ and $2$,
the conductivity of our model diverges. In order to understand the
nature of this divergence we define the (microcanonical) conductivity
of the finite system of size $N$ as
\begin{equation}
  \label{eq:6}
  \kappa_N^{1,1} (T) =   \frac 1{2 T^2 t_N} \frac 1{N^{d}}\ME
  \left(\left[\sum_{\bx \in {\mathbb Z}_N^d} J_{\bx,\bx+\be_1}([0,t_N])\right]^2 \right) 
\end{equation}
where $t_N= N/v_s$ with $v_s = \lim_{\bk\to 0} |\partial_{k_1}
\omega(\bk)| = 2\alpha^{1/2}$ the sound velocity. 
This definition of the conductivity of the finite system is motivated
by the following consideration: 
$\nabla_{k}\omega(\bk)$ is the group velocity of
the $\bk$-mode waves, and typically $v_s$ is an upper bound for these
velocities. Consequently $t_N$ is the typical time a
 low $\bk$ (acoustic) mode takes to cross around the system once. One defines similarly ${\tilde \kappa}_N^{1,1} (T)$ by
 \begin{equation}
  \label{eq:6c}
   \tilde\kappa_N^{1,1}(T) =   \frac 1{2 T^2 t_N} \frac 1{N^{d}}\CE
  \left(\left[\sum_{\bx \in {\mathbb Z}_N^d} J_{\bx,\bx+\be_1}([0,t_N])\right]^2 \right)
\end{equation}

 We conjecture  that $\kappa_N$ (resp. ${\tilde \kappa}_N$) has the same asymptotic behavior as the conductivity
 defined in the non-equilibrium stationary state on the open system
 with thermostats at the boundary at different temperature, as
 defined in eg. \cite{bo, blr, rll}.

With these definitions we have the following theorem:  
\begin{theo}
  \label{th-ldharm}
In the harmonic case, if $W= 0$:
\begin{enumerate}
\item $\kappa_N \sim N^{1/2}$ if $d=1$,
\item $\kappa_N \sim \log N$ if $d=2$.
\end{enumerate}
In all other cases $\kappa_N$ is bounded in $N$ and converges to
$\kappa$. \\
Same results are valid for ${\tilde \kappa}_N$.
\end{theo}

In fact we show that, in the harmonic case, we have
\begin{equation}
\lim_{N \to \infty} \cfrac{{\tilde \kappa}_{N}^{1,1} (T)}{\kappa_N^{1,1} (T)}=1
\end{equation}
This is a consequence of equation (\ref{eq:C11}) that one can easily check is also valid if the microcanonical measure is replaced by the canonical measure.\\

In the anharmonic case we cannot prove the existence of neither
$ \tilde \kappa^{1,1}(T)$ nor $\kappa^{1,1} (\mathcal E)$, but we can
establish upper bounds for the canonical version of the finite size
Green-Kubo formula (\ref{eq:6c}).  
Extra assumptions on the potentials $V$ and $W$ assuring a uniform
control on the canonical static correlations (see
(\ref{eq:deco2}-\ref{eq:deco1})) have to be done. In the unpinned case
$W=0$, (\ref{eq:deco1}) is valid as soon as $V$ is strictly convex. In
the pinned case $W>0$, (\ref{eq:deco2}) is ``morally'' valid as soon as
the infinite volume Gibbs measure is unique.  Exact assumptions are
given in \cite{BH}, theorem 3.1 and theorem 3.2. In the sequel,
"general anharmonic case" will refer to potentials $V$ and $W$ such
that $(\ref{eq:deco2})$ (or (\ref{eq:deco1})) is valid.

\begin{theo}
  \label{th-anharm}
Consider the general anharmonic case. There exists a constant $C$
(depending on the temperature $T$) such that 
 \begin{itemize}
\item For $d\ge 3$, 
  \begin{enumerate}\item 
    either $W>0$ is general
  \item or if $W=0$ and $0< c_-\le V_j''\le C_+< \infty$ for any
    $j$, 
  \end{enumerate}
 then 
$$
 \tilde\kappa^{1,1}_N (T) \le C.
$$

 \item For $d=2$, if $W=0$ and $0< c_-\le V_j''\le C_+< \infty$ for
   any $j$, then    
 $$
 \tilde\kappa^{1,1}_N (T) \le C(\log N)^2.
 $$
 \item For $d=1$, if $W=0$ and $0< c_-\le V''\le C_+< \infty$, then 
 $$
 \tilde\kappa^{1,1}_N (T) \le C {\sqrt N}.
 $$
\item
Moreover, in any dimension, if $V_j$ are quadratic and $W>0$ is
general then ${\tilde \kappa}^{1,1}_N (T)\leq C$.  
\end{itemize}
\end{theo}

The proof of this statement is in section
\ref{sec:anharm-case:-bounds}.\\

We now relate the definition of the Green-Kubo (\ref{eq:gk0}) and
\eqref{eq:gcc}  to the
variance of the energy-energy correlations function
(\ref{eq:variance}). \\ 

Consider the infinite volume dynamics on ${\mathbb Z}^d$ under the infinite volume canonical Gibbs measure with temperature $T>0$. The expectation is denoted by ${\E}_{\scriptscriptstyle T}$. Fix $t>0$ and assume that the following sum makes sense
\begin{equation}
D^{i,j}_{\scriptscriptstyle T} (t)=\sum_{\bx \in {\mathbb Z}^d} \bx^i
\bx^j \E_{\scriptscriptstyle T} \left[ \, ({\cE}_\bx(t) -T)(\cE_0(0) -T)\,
\right] = \sum_{\bx \in {\mathbb Z}^d} \bx^i
\bx^j S(\bx, t).
\end{equation}

If $\bx \neq 0$, by space and time invariance of the dynamics, we have
\begin{equation}
\E_{\scriptscriptstyle T} \left[({\cE}_{\bx} (t) -T)({\cE}_0 (0)
  -T)\right] =-\cfrac{1}{2} \E_{\scriptscriptstyle T}\left[ \,
  ({\cE}_\bx (t) - {\cE}_\bx (0)) \, ( {\cE}_0 (t) - {\cE}_0 (0))\,
\right]  
\end{equation}
By definition of the current, we have for any $\by \in {\mathbb Z}^d$:
\begin{equation}
{\cE}_\by (t) -{\cE}_\by (0)= \sum_{k=1}^d \left(J_{\by-\be_k,\by}
  ([0,t]) - J_{\by,\by+\be_k} ([0,t])\right) 
\end{equation}
By two discrete integration by parts one obtains
\begin{equation}
D^{i,j}_{\scriptscriptstyle T} (t)= \sum_{x \in {\mathbb Z}^d} \E_T
\left[ J_{\bx,\bx+\be_i} ([0,t]) \, J_{0,\be_j} ([0,t])\right] 
\end{equation}
so that the thermal conductivity is equal to the space-time
correlations of the total current 
\begin{equation}
\kappa^{i,j} (T)=\delta_{0} (i-j) \,\lim_{t \to \infty} \cfrac{1}{2T^2
  t} \sum_{x \in {\mathbb Z}^d}\E_{\scriptscriptstyle T} \left[
  J_{\bx,\bx+\be_i} ([0,t]) \, J_{0,\be_j} ([0,t])\right] 
\end{equation}

Of course this derivation is only formal even for fixed time $t
>0$. The problem is to define the infinite volume dynamics and to show
$S(\bx,t)$ has a sufficiently fast decay in $\bx$. For the purely
Hamiltonian dynamics, it is a challenging problem. For the stochastic
dynamics it seems less difficult but remains technical. To avoid these
difficulties we adopt a finite volume limit procedure starting from
(\ref{in:gk}). This explains the definitions (\ref{eq:gk0}) and
(\ref{eq:gcc}).\\ 

Consider now the closed dynamics on ${\mathbb Z}_N^d$ starting from
the microcanonical state. The rest of the section is devoted to the
proof of the following formula 
\begin{equation}
  \label{eq:5}
  \begin{split}
  &  \frac 1{2 T^2 t} \frac 1{N^{d}}\ME \left(\left[\sum_{\bx
        \in \To_N^d} 
        J_{\bx,\bx+\be_1}([0,t])\right]^2 \right) \\
    &= (2 T^2 t N^d)^{-1} \ME \left(\left[\sum_{\bx \in \To_N^d} \int_0^t
        j^a_{\bx,\bx+\be_1}(s) ds\right]^2 \right) + \frac{\gamma}d 
    + \frac{O_N}{N^d}
  \end{split}
\end{equation}
and an identical formula in the canonical case (with $\ME$ substituted
by $\mathbb E_{N,T}$).

The term $\gamma/d$ in (\ref{eq:5}) is the direct contribution of the
stochastic dynamics to the thermal conductivity.
In the microcaconical case we actually prove
that is is equal to $\gamma/d$ only for the harmonic case. 
A complete proof of (\ref{eq:5}) for anharmonic interaction demands 
an extension of the equivalence of ensembles estimates proven in
section \ref{sec:equivalence}. In the grancanonical case this problem
does not appear.

Starting in the microcanonical case,
remark that the first term on the RHS of (\ref{eq:5}) can be written as
\begin{equation}
\label{eq:daw}
\begin{split}
&(2 T^2 t N^d)^{-1}{\ME}\left(\left[\sum_{\bx\in \To_N^d}\int_0^t
    j^a_{\bx,\bx+\be_1}(s) ds\right]^2 \right)\\ 
&=\cfrac{1}{T^2} \int_0^{\infty}
\left(1-\cfrac{s}{t}\right)^+ \sum_{x \in {\mathbb Z}_N^d} {\ME}
\left( j^a_{\bx,\bx+\be_1}(s) j^a_{0,\be_1} (0)\right) ds 
\end{split}
\end{equation}

If $\gamma=0$, which corresponds to the purely Hamiltonian system, as
$N$ and then $t$ goes to infinity, and if one can prove that the
current-current correlation function has a sufficiently fast decay,
then one recovers the usual Green-Kubo formula (\ref{in:gk}).  

To prove (\ref{eq:daw}) one uses space and time translation invariance
of the dynamics 
\begin{eqnarray*}
&&(2 T^2 t N^d)^{-1}{\ME}\left(\left[\sum_\bx \int_0^t
    j^a_{\bx,\bx+\be_1}(s) ds\right]^2 \right)\\ 
&=&(2 T^2 t N^d)^{-1}\sum_{\bx,\by} \int_0^t ds \int_{0}^t du\; {\ME}
\left(j^a_{\bx,\bx+\be_1}(s) \, j^{a}_{\by,\by +\be_1} (u)\right)\\ 
&=& (T^2 t N^d)^{-1} \sum_{\bx,\by} \int_0^t ds \int_{0}^s du\; {\ME}
\left(j^a_{\bx,\bx+\be_1}(s) \, j^{a}_{\by,\by+\be_1} (u)\right)\\ 
&=& (T^2 t N^d)^{-1} \sum_{\bx,\by} \int_0^t ds \int_{0}^s du\; {\ME}
\left(j^a_{\bx-\by,\bx-\by+\be_1}(s-u) \, j^{a}_{0,\be_1} (0)\right)\\ 
&=& \cfrac{1}{T^2} \int_0^{\infty}
\left(1-\cfrac{s}{t}\right)^+ \sum_{x} {\ME} \left(
  j^a_{\bx,\bx+\be_1}(s) j^a_{0,\be_1}(0) \right) ds 
\end{eqnarray*}

We now give the proof of (\ref{eq:5}). Because of the periodic
boundary conditions, since $j^s$ if a 
\emph{gradient} (cf. (\ref{eq:3})), the corresponding  terms cancel,
and we can write 
\begin{equation}
\label{decomp0}
\begin{array}{lcl}
     \sum_\bx J_{\bx,\bx+\be_1}([0,t]) &=& \int_0^t \sum_\bx j^a_{\bx,
       \bx+\be_1}(s) \; ds + \sum_\bx M_{\bx,\bx+\be_1}(t)\\
       &=& \int_0^t {\frak J}_{\be_1}(s) \; ds + {\frak M}_{\be_1}(t)
\end{array}
\end{equation}
so that 
\begin{equation}
\label{eq:2005}
\begin{split}
&(t N^d)^{-1}{\ME}\left(\left[\sum_\bx J_{\bx,\bx+\be_1}([0,t])\right]^2 \right)\\
&= (tN^d)^{-1}{\ME}\left(\left[\int_0^t {\frak J}_{\be_1}(s)
    ds\right]^2 \right)+(tN^{d})^{-1}{\ME}\left({\frak M}^2_{\be_1}(t)\right)\\ 
&\qquad +2(tN^d)^{-1}{\ME}\left(\left[\int_0^t {\frak J}_{\be_1}(s) \;
    ds\right]{\frak  M}_{\be_1} (t) \right)  
\end{split}
\end{equation}

The third term on the RHS of (\ref{eq:2005}) is shown to be zero by a
time reversal argument and the second term on the RHS of
(\ref{eq:2005}) gives in the limit a contribution equal to
$\gamma/d$. 

To see the first claim let  us denote by $\{\omega (s)\}_{0\leq s \leq
  t}$ the process \hfill \\
$\{(\bp_\bx (s), \bq_\bx (s)) ;\ \bx \in {\mathbb
  Z}_N^d,\ 0 \leq s \leq t\}$ 
arising in (\ref{eq:sde}) or in (\ref{eq:sde1}) for the
one-dimensional case. The reversed process $\{\omega^*_s\}_{0 \leq s
  \leq t}$ is defined as $\omega^*_s = \omega_{t-s}$. Under the
microcanonical measure, the time reversed process is still Markov with
generator $-A + \gamma S$. The total current $J_t
(\omega_\cdot)=\sum_{\bx} J_{\bx, \bx +\be_1} ([0,t])$ is a
functional of $\{\omega_s\}_{0\leq s \leq t}$. By
(\ref{eq:sde}-\ref{eq:sde1}), we have in fact that $J_t(\cdot)$ is an
anti-symmetric functional of $\{\omega_s\}_{0\leq s \leq t}$, meaning 
\begin{equation}
\label{eq:rev}
J_t (\{\omega^*_s\}_{0 \leq s \leq t})= -J_t(\{\omega_s\}_{0 \leq s \leq t})
\end{equation} 
In fact, similarly to (\ref{eq:tc}), we have
\begin{equation}
J_s (\omega^*_\cdot)=\int_0^s {({\frak J}_{\be_1})}^{*}(\omega^*(v))dv +{\frak M}_{\be_1}^* (s),
\quad 0 \leq s \leq t 
\end{equation}
where $({\frak M}_{\be_1}^* (s))_{0 \leq s \leq t}$ is a martingale with respect to
the natural filtration of $(\omega^*_s)_{0\leq s \leq t}$ and
$({\frak J}_{\be_1})^*=\sum_\bx (j^{a})^*_{\bx, \bx + \be_1}$ is equal to $-{\frak J}_{\be_1}=
-\sum_{\bx} j^a_{\bx, \bx + \be_1}$.   

We have then by time reversal
\begin{equation}
\label{eq:tr}
\begin{split}
{\ME}[J_t (\omega_\cdot) {\frak J}_{\be_1} (\omega (t))] &= -{\ME} [J_t
(\omega^*_\cdot) {\frak J}_{\be_1} (\omega^* (0))] \\ 
&=-{\ME} \left[ \left(\int_0^t ({\frak J}_{\be_1})^*(\omega^*(s))ds + {\frak M}^*
    (t)\right){{\frak J}_{\be_1}}(\omega^* (0))\right]\\ 
&=-{\ME} \left[ \left(\int_0^t {{\frak J}_{\be_1}}^*(\omega^*(s))ds\right) {\frak J}_{\be_1}
  (\omega^*(0))\right]
\end{split}
\end{equation}
where the last equality follows from the martingale property of
${\frak M}^*$. Recall now that $({\frak J}_{\be_1})^*=-{\frak J}_{\be_1}$. By variables change $s \to t-s$
in the time integral, we get 
\begin{equation}
\label{eq:tr2}
{\ME}[J_t (\omega_\cdot) {\frak J}_{\be_1} (\omega (t))]={\ME} \left[
  \left(\int_0^t {\frak J}_{\be_1} (\omega(s))ds\right) {\frak J}_{\be_1} (\omega (t))\right] 
\end{equation}

It follows that 
\begin{equation}
\begin{split}
  {\ME}\left[\left(\int_0^t {\frak J}_{\be_1}
    (\omega(s))ds\right){\frak M}_{\be_1}(t)\right]={\ME}\left[\int_0^t {\frak J}_{\be_1}
  (\omega(s)) {\frak M}_{\be_1}(s)ds\right]\\ 
   =\int_0^t  ds\; {\ME}\left[ {\frak J}_{\be_1} (\omega(s)) \left (J_s (\omega_{\cdot})
    - \int_0^s {\frak J}_{\be_1} (\omega (v)) dv\right)\right]
    = 0
\end{split}
\end{equation}

For the second term on the RHS of (\ref{eq:2005}) we have 
\begin{equation*}
   \begin{split}
  (tN^{d})^{-1}\ME\left({\frak M}^2_{\be_1} (t)\right) =&\frac{
    \gamma}{(d-1) N^{d}} 
 \sum_\bx \sum_{i,j} \left< \left(X^{i,j}_{\bx, \bx+\be_1}
     (\bp_\bx^2/2) \right)^2\right>_{\scriptscriptstyle{N,\mathcal E}} \\ 
  =& \frac{ \gamma}{(d-1)N^{d}} \sum_\bx \sum_{i\neq j} \left< 
    \left( p_\bx^j p_{\bx + \be_1}^i - p_\bx^i p_{\bx + \be_1}^j
    \right)^2\right>_{\scriptscriptstyle{N,\mathcal E}}  \\
  =& \frac{2 \gamma}{(d-1)N^{d}} \sum_\bx \sum_{i\neq j} \left<  (p_\bx^j
    p_{\bx + \be_1}^i)^2  \right>_{\scriptscriptstyle{N,\mathcal E}}
  \\
 &  - \frac{2 \gamma}{(d-1)N^{d}} \sum_\bx 
  \sum_{i\neq j} \left<  (\bp_{\bx}^i \bp_{\bx + \be_1}^i \bp_\bx^j
    \bp_{\bx + \be_1}^j)  \right>_{\scriptscriptstyle{N,\mathcal E}} 
\end{split}
\end{equation*}
Thanks to the equivalence of ensembles (cf. lemma
\ref{lem:equiv-ens}), this last quantity is equal to    
\begin{equation}
\label{eq:f28}
2 \gamma \frac{ T^2}{d} + N^{-d} O_N
\end{equation}
where $O_N$ remains bounded as $N\to \infty$. The calculation in $d=1$
is similar. The contribution of the martingale term for the
conductivity is hence $\gamma/d$ and we have shown (\ref{eq:5}). 
Notice this is the only point where we have used the equivalence of
ensembles results of section \ref{sec:equivalence} that we have proven   
only in the harmonic case. We conjecture these are true also for the
anharmonic cases.

Observe that all the arguments above between (\ref{decomp0}) and
(\ref{eq:f28}) apply directly also to the canonical definition of the
Green-Kubo but without the small error in $N$ (because for the
canonical measure momentums $\bp_\bx$ are independently distributed
and the equivalence of ensembles approximations are in fact
equalities). Therefore we have the similar formula to (\ref{eq:5}): 
\begin{equation}\label{eq:5c}
   \begin{split}
   & \frac 1{2 T^2 t} \sum_\bx \CE \left(
        J_{\bx,\bx+\be_1}([0,t]) J_{0,\be_1}([0,t]) \right)\\
   & = (2 T^2 N^d t)^{-1} \CE\left(\left[\sum_\bx \int_0^t
        j^a_{\bx,\bx+\be_1}(s) ds\right]^2 \right) + \frac{\gamma}d
  \end{split}
\end{equation}

In the next sections
we will consider the $(\alpha,\nu)$-harmonic case and  we will compute
explicitly the limit  (as $N \to \infty$ and then $t\to \infty$) of
the two first term on the RHS of (\ref{eq:2005}).  

%%%%%%%%%%%%%%%%%%%%%%%%%%%%%%%%%%%%%%%%%%%%%%%%%%%%%%%%%%%%%%%%%%%%%%%
%%%%%%%%%%%%%%%%%%%%%%%%%%%%%%%%%%%%%%%%%%%%%%%%%%%%%%%%%%%%%%%%%%%%%%%
%%%%%%%%%%%%%%%%%%%%     correlation   %%%%%%%%%%%%%%%%%%%%%%%%%%%%%%%%
%%%%%%%%%%%%%%%%%%%%                   %%%%%%%%%%%%%%%%%%%%%%%%%%%%%%%%
%%%%%%%%%%%%%%%%%%%%%%%%%%%%%%%%%%%%%%%%%%%%%%%%%%%%%%%%%%%%%%%%%%%%%%%
%%%%%%%%%%%%%%%%%%%%%%%%%%%%%%%%%%%%%%%%%%%%%%%%%%%%%%%%%%%%%%%%%%%%%%%

\section{ Correlation function of the energy current in the harmonic case}
\label{sec:cfe}

% \marginpar{\emph{Do we need rally to do this firther conditioning on the center of mass?}}
% The microcanonical measure  is usually defined as the uniform measure
% on the energy surface defined by ${\mathcal H}_N =N^d e$. Our dynamics
% conserve also $\left(\sum_x \bp_\bx\right)^2 + \nu \left( \sum_{\bx}
%   \bq_\bx\right)^2$. Notice that the dynamics is invariant under the
% change of coordinates $\bp^{'}_{\bx}=\bp_\bx - \sum_{\by} \bp_{\by}$ and
% $\bq^{'}_{\bx}=\bq_{\bx} -\sum_{\by} \bq_{\by}$. Consequently, without
% any lost of generality, we can fix $\sum_{\bx} \bp_{\bx} =0$ and
% $\sum_{\bx} \bq_{\bx} =0$. So in  the following we define as
% microcanonical measure  the uniform probability measure on the 
% $(N^{2d} - 2d -1)$-dimensional sphere
% \begin{equation*}
%   \left\{ {\mathcal H}_N =N^d e;\; \sum_{\bx} \bp_{\bx} =0;\;
%     \sum_{\bx} \bq_{\bx} =0 \right\} 
% \end{equation*}
% This is the unique invariant measure for the dynamics on this surface.
% We denote by $<\cdot>_{\scriptscriptstyle{N,e}}$ (sometimes, we will omit the subscript $N$) the expectation with respect to the microcanonical measure .  
We consider the $(\alpha,\nu)$-harmonic case (\ref{eq:harm}). We recall that $\frak J_{\be_1} = \sum_\bx j_{\bx,\bx+\be_1}$. Because of the periodic boundary conditions, and being
 $j^s_{\bx, \bx+\be_1}$ a spatial gradient (cf. (\ref{eq:3})), we have
 that $\frak J_{\be_1} = \sum_\bx j^a_{\bx, \bx+\be_1}$. 
We are interested in the decay of the
correlation function:
\begin{equation}
  \label{eq:8}
  C_{1,1} (t) = \lim_{N\to \infty} \frac 1{N^d} \ME(\frak J_{\be_1}(t)\frak
  J_{\be_1}(0)) =  \lim_{N\to \infty} \sum_\bx  \ME( j^a_{0,
    \be_1}(0)  j^a_{\bx, \bx + \be_1}(t))
\end{equation}
where $\ME$ is the expectation starting with the
microcanonical distribution defined above.

 For $\lambda > 0$, let $u_{\lambda,N}$ be the solution of the Poisson equation 
$$\lambda u_{\lambda,N} - Lu_{\lambda,N} = -\sum_\bx j^a_{\bx, \bx + \be_1}$$  
given explicitely in lemma \ref{lem:u} of section \ref{harm2}.  By lemma
\ref{lem:inversion}, 
we can write   
the Laplace transform of $C_{1,1}(t)$ as
\begin{equation}
  \label{eq:110}
  \int_0^\infty dt e^{-\lambda t} C_{1,1}(t) \; dt= \lim_{N \to
    \infty} \left<j^a_{0,\be_1}u_{\lambda,N}\right>_{\scriptscriptstyle{N,\mathcal E}}  
\end{equation}

Substituting in (\ref{eq:110}) the explicit form of $u_{\lambda,N}$ given in
lemma \ref{lem:u}, we have: 
\begin{equation}
\label{eq:sym1}
\begin{split}
 -  \left<j^a_{0,\be_1}
   u_{\lambda,N}\right>_{\scriptscriptstyle{N, \mathcal E}}
 = \frac{\alpha^2}{2 
   \gamma} \sum_{\bx,\by} g_{\lambda,N}(\bx-\by) \left<(\bq_{\be_1} - \bq_0)
   \cdot (\bp_{\be_1} + \bp_0) (\bp_\bx \cdot \bq_\by)
 \right>_{\scriptscriptstyle{N, \mathcal E}}\\  
=\frac{\alpha^2}{2 \gamma} \sum_{\bx,\by} g_{\lambda,N}(\bx-\by)
\left<(\bq_{\be_1}\cdot \bp_0 - \bq_0 \cdot \bp_{\be_1}) (\bp_\bx
  \cdot \bq_\by) \right>_{\scriptscriptstyle{N,\mathcal E}} \\
+  \frac{\alpha^2 }{2 \gamma} \sum_{\bx,\by} g_{\lambda,N}(\bx-\by)
\left<(\bq_{\be_1}\cdot \bp_{\be_1} - \bq_0 \cdot \bp_0) (\bp_\bx
  \cdot \bq_\by) \right>_{\scriptscriptstyle{N,\mathcal E}}
\end{split}
\end{equation}
Observe that the last term on the RHS of (\ref{eq:sym1}) is null by
the translation invariance property.
So we have (using again the translation invariance and the
antisymmetry of $g_{\lambda,N}$)
\begin{equation*}
  \begin{split}
    -  \left<j^a_{0,\be_1} u_{\lambda,N}\right>_{\scriptscriptstyle{N,e}} = 
     \frac{\alpha^2}{2 \gamma} \sum_{\bx,\by} g_{\lambda,N}(\bx-\by) 
     \left<(\bq_{\be_1}- \bq_{-\be_1}) 
         \cdot \bp_0) (\bp_\bx\cdot \bq_\by)% \\
%          = -  \frac{\alpha^2 d}{2 \gamma e^2} \sum_{\bx,\by} g_N(\by-\bx) 
%      \left<(\bq_{\be_1}- \bq_{-\be_1}) 
%          \cdot \bp_0) (\bp_\bx\cdot \bq_\by)
        \right>_{\scriptscriptstyle{N,e}}
  \end{split}
\end{equation*}

Define 
$$
K_N (\bq)= N^d \mathcal E -\cfrac{1}{2} \sum_{\bx} \bq_{\bx}\cdot (\nu I
-\alpha \Delta) \bq_{\bx}
$$
In the unpinned case $\nu=0$, conditionally to the positions configuration $\bq$,
the law of $\bp$ is $\mu_{\bq}=\mu_{\sqrt{2K_N (\bq)}}^{N^d}$ (defined
in lemma \ref{lem:sym2}), meaning the uniform measure on the surface 
% \marginpar{\emph{In the pinned case \\we should not fix\\ the total
%     momentum!}}
$$
\left\{(\bp_\bx)_{\bx \in {\mathbb Z}_N^d}; \quad
  \cfrac{1}{2}\sum_{\bx} \bp_\bx^2 = K_N (\bq); \quad \sum_{\bx}
  \bp_{\bx} =0  \right\}
$$ 
  
By using properties (i),(ii) and (iii) of lemma \ref{lem:sym2}, one
has for $\bx\neq 0$, 
\begin{equation}
\begin{split}
\left<((\bq_{\be_1}- \bq_{\be_1})\cdot \bp_0)(\bp_{\bx} \cdot \bq_{\by})\right>_{\scriptscriptstyle{N,\mathcal E}} =
\sum_{i,j}\left< \mu_{\bq} \left(\bp_0^i \bp_{\bx}^j\right)(\bq_{\be_1}^i-
  \bq_{-\be_1}^i) \bq_{\by}^j \right>_{\scriptscriptstyle{N,\mathcal E}}\\  
= \sum_{i}\left< \mu_{\bq} \left(\bp_0^i 
    \bp_{\bx}^i\right)(\bq_{\be_1}^i-\bq_{-\be_1}^i) \bq_{\by}^i
\right>_{\scriptscriptstyle{N,\mathcal E}}\\  
=-\sum_{i=1}^{d} \left<\cfrac{2K_N (\bq)}{dN^d (N^d -1)}(\bq_{\be_1}^i-
  \bq_{-\be_1}^i)  \bq_{\by}^i\right>_{\scriptscriptstyle{N,\mathcal E}}\\ 
= -\cfrac{1}{N^d -1} \sum_{i=1}^d \left< (\bp_{0}^{i})^2
  (\bq_{\be_1}^i- \bq_{-\be_1}^i) \bq_{\by}^i\right>_{\scriptscriptstyle{N,\mathcal E}} 
\end{split}
\end{equation} 
For $\bx =0$, one gets 
\begin{equation}
\begin{split}
\left<((\bq_{\be_1}- \bq_{\be_1})\cdot \bp_0)(\bp_{0} \cdot
  \bq_{\by})\right>_{\scriptscriptstyle{N,\mathcal E}} = \sum_{i,j}\left<
  \mu_{\bq} \left(\bp_0^i 
    \bp_{0}^j\right) (\bq_{\be_1}^i- \bq_{-\be_1}^i) \bq_{\by}^j
\right>_{\scriptscriptstyle{N,\mathcal E}}\\  
= \sum_{i=1}^d \left< \mu_{\bq} \left(\bp_0^i \bp_{0}^i\right)
  (\bq_{\be_1}^i- \bq_{-\be_1}^i) \bq_{\by}^i \right>_{\scriptscriptstyle{N,\mathcal E}}\\
=\sum_{i=1}^d \left<\left(\bp_{0}^i\right)^2  (\bq_{\be_1}^i-
  \bq_{-\be_1}^i)  \bq_{\by}^i \right>_{\scriptscriptstyle{N,\mathcal E}}
\end{split}
\end{equation}

In the pinned case $\nu>0$, conditionally to the positions configuration $\bq$,
the law of $\bp$ is $\lambda_{\bq}=\lambda_{\sqrt{2K_N (\bq)}}^{N^d}$ (defined
in lemma \ref{lem:sym}), meaning the uniform measure on the surface 
$$ \left\{(\bp_\bx)_{\bx \in {\mathbb Z}_N^d}; \quad
  \cfrac{1}{2}\sum_{\bx} \bp_\bx^2 = K_N (\bq) \right\} $$ 
We proceed in a similar way and we observe that if $\bx \neq 0$, $\lambda_{\bq} (\bp_0^i \bp_\bx^i)=0$.(cf. ii) of lemma \ref{lem:sym})

 Since $g_{\lambda,N}$ is antisymmetric (see (\ref{eq:g}-\ref{eq:g1d})) and such that $\sum_{\bz} g_{\lambda,N} (\bz) =0$, one obtains easily in both cases (pinned and unpinned)

\begin{equation}
\label{eq:sym2005}
\begin{split}
 &-  \left<j^a_{0,\be_1} u_{\lambda,N}\right>_{\scriptscriptstyle{N,e}}=
 - \frac{\alpha^2 }{2 \gamma} \sum_{\by} g_{\lambda,N}(\by)  \sum_i 
\left<\left(p_{0}^i\right)^2  (q_{\be_1}^i- q_{-\be_1}^i)  q_{\by}^i
\right>_{\scriptscriptstyle{N,\mathcal E}}\\
&+ \frac{\alpha^2}{2 \gamma}\frac {\bf{1}_{\nu=0}}{N^d -1} \sum_{\bx\neq 0,\by}
g_{\lambda,N}(\by-\bx) \sum_i 
\left<\left(p_{0}^i\right)^2  (q_{\be_1}^i- q_{-\be_1}^i)  q_{\by}^i \right>_{\scriptscriptstyle{N,\mathcal E}}
\\ 
& = -\left( 1+ \cfrac{{\bf 1}_{\nu=0}}{N^d -1}\right) \frac {\alpha^2
}{2 \gamma}  
\sum_{\by} g_{\lambda,N} (\by) \sum_i 
\left<\left(p_{0}^i\right)^2  (q_{\be_1}^i- q_{-\be_1}^i)  q_{\by}^i
\right>_{\scriptscriptstyle{N,\mathcal E}} . 
\end{split}
\end{equation}

%We work out first the case $d\ge 2$ or $\nu >0$.

Let $\Gamma_N (\bx)$, $\bx \in {\To_N^d}$, be the unique solution of
\begin{equation}
\label{eq:gamma}
\left(\nu I - \alpha \Delta \right) \Gamma_N = \delta_{\be_1} - \delta_{-\be_1}
\end{equation}
such that $\sum_{\bx \in {\To}_N^d} \Gamma_N (\bx) =0$.

By (iii) of lemma \ref{lem:equiv-ens} and (\ref{eq:16}), we have 
\begin{equation}
\label{deco}
\begin{split}
   \left| - \left<j^a_{0,\be_1}
       u_{\lambda,N}\right>_{\scriptscriptstyle{N,\mathcal E}}    
   - \left(1+ \cfrac{{\bf 1}_{\nu=0}}{N^d -1} \right)\frac{\alpha^2
     \mathcal E^2}{2 \gamma d} \sum_{\by}  g_{\lambda,N}(\by)
   \Gamma_N(\by)\right| \\ 
  \le \frac{C\log N}{N^{d}}
  \sum_{\by} \left|g_{\lambda,N}(\by)\right| \le   \frac{C \log N}{N^{d/2}}
 \left(\sum_{\bx} ( g_{\lambda,N} (\bx))^2 \right)^{1/2} 
 \le \frac{C'\log N}{\lambda N^{d/2}} 
\end{split}
\end{equation}
%It is well known (cf. (\ref{eq:satd})) that $\Gamma_N(0) \sim \log N$ if $d=2$ and $\nu =0$, while it stays bounded in the cases $d\geq 3$ or $\nu>0$. If $d=1, \nu =0$, $\Gamma_N (0)$ is of order $N$. 
%This is the reason we need to consider this case separately. 

Hence the last term of (\ref{deco}) goes to $0$. 

Taking the limit as $N\to \infty$ we obtain (see (\ref{eq:conv-RS}))
\begin{equation}\label{eq:ggg}
  \begin{split}
    \int_0^\infty  e^{-\lambda t} C_{1,1}(t) \; dt =
     \frac {\alpha^2
      \mathcal E^2}{2 d \gamma} \sum_\bz g_\lambda(\bz) \Gamma(z)
  \end{split}
\end{equation}
where $g_\lambda$ are solutions of the same equations as
$g_{\lambda,N}$ but on $\mathbb Z^d$ and $\Gamma$ is the solution of
the same equation as $\Gamma_N$ but on $\Z^d$. 

Using Parseval relation and the explicit form of the Fourier transform of $g_\lambda$  (cf. (\ref{eq:gg2})) and $\Gamma$, one gets the following formula for the Laplace transform of $C_{1,1} (t)$ for $d\geq 2$:
\begin{equation}
\label{eq:lt-C11}
\cfrac{\alpha^2 \mathcal E^2}{d} \int_{[0,1]^d} d \bk
\left(\cfrac{\sin^{2}(2\pi \bk^1)}{\nu + 4\alpha \sum_{j=1}^d \sin^{2}
    (\pi \bk^j)}\right)\cfrac{1}{\lambda + 8\gamma \sum_{j=1}^d
  \sin^{2} (\pi \bk^j)} 
\end{equation}

By injectivity of Laplace tranform, $C_{1,1} (t)$ is given by:
\begin{equation}
\label{eq:C11-d}
C_{1,1} (t) =\cfrac{\alpha^2 \mathcal E^2}{d} \int_{[0,1]^d} d \bk \left(\cfrac{\sin^{2}(2\pi \bk^1)}{\nu + 4\alpha \sum_{j=1}^d \sin^{2} (\pi \bk^j)}\right) \exp\left\{\ -8 \gamma t \sum_{j=1}^d \sin^{2} (\pi \bk^j)\right\}
\end{equation}

For the one dimensional case, the equation for $g_{\lambda,N}$ (resp. $g_\lambda$) is different (see (\ref{eq:gg1}) ) and we get the following integral representation of the correlation function of the energy current:

\begin{equation}
\label{eq:C11-1}
C_{1,1} (t)= \alpha \mathcal E^2 \int_{0}^1 dk \cos^2 (\pi k)
\exp\left\{ -\cfrac{4 \gamma t}{3} \sin^{2} (\pi k) (1 + 2 \cos^2 (\pi
  k) \right\} 
\end{equation}

%In the 1-dimensional case, if $\nu = 0$ we have to proceed a bit differently. 
%Let $G_{\lambda,N}(z)$ be the function on $\mathbb Z_N$ solution of the
%equation (\ref{eq:15}) for $d=1$.

%We shows in section \ref{harm2} that $G_{\lambda, N}$ behaves basically as the
%Green function of $(-\Delta)$  on $\mathbb Z_N$ as $N\to \infty$ and
%that  $g_{\lambda,N} (z) =  \left( G_{\lambda,N}(z+1) -  G_{\lambda,N}(z-1) \right)$.
% We use the notation $r_x = q_{x+1} - q_x$.  Translation invariance and symmetry properties of $G_{\lambda,N}$ give 
%\begin{equation}
 % \label{eq:8-new}
 % \begin{split}
  %  -\sum_{y} g_{\lambda,N}(y) &\left< p_0^2 (q_{1}- q_{-1}) \; q_y \right>_{\scriptscriptstyle{N,e}}\\
   %  &= -  \sum_{y}  G_{\lambda,N}(y) \left< p_0^2 (q_{1}- q_{-1}) (q_{y-1}
   %   -q_{y+1})  \right>_{\scriptscriptstyle{N,e}} \\
   % & =  \sum_{y}  G_{\lambda,N}(y) \left< p_0^2 (r_0 + r_{-1}) (r_{y} +
   %    r_{y-1})   \right>_{\scriptscriptstyle{N,e}}
 % \end{split}
%\end{equation}

In any dimension, we have the following unified formula for $C_{1,1} (t)$ 
\begin{equation}
\label{eq:C11}
C_{1,1} (t) = \cfrac{\mathcal E^2}{4\pi^2 d}\int_{[0,1]^d}
(\partial_{\bk^1} \omega (\bk))^2 e^{-t\gamma \psi (\bk)} d\bk 
\end{equation}
where  $\omega(\bk)$ is defined by (\ref{eq:dr}) and $\psi (\bk)$ by 
(\ref{eq:psi}).
% \begin{equation}
% \label{eq:dr}
% \omega (\bk) = (\nu + 4\alpha\sum_{j=1}^{d} \sin^{2} (\pi \bk^j))^{1/2}
% \end{equation}
% is the dispertion relation of the system, and 
% \begin{equation}
% \label{eq:psi}
% \psi (\bk) = 
% \begin{cases}
% 8\sum_{j=1}^{d} \sin^{2} (\pi \bk^j), \quad \mbox{ if } \quad d\geq 2\\
% 4/3 \ \sin^2 (\pi \bk) (1+2\cos^2 (\pi \bk)), \quad \mbox{ if } \quad
% d=1 .
% \end{cases}
% \end{equation}
Observe that the same formula holds if we replace $\ME$ by $\CE$. 
In this last case, the situation is simpler since we do not need 
equivalence of ensembles.\\

Standard analysis shows the behavior of $C_{1,1}(t)$ as $t$ goes to
infinity is governed by the behavior of the function $(\partial_{\bk^1}
\omega (\bk))^2$ and $\psi (\bk)$ around the minimal value of $\psi$ which
is $0$. In fact, $\psi (\bk)=0$ if and only if $\bk=0$ or $\bk
=(1,\ldots,1)$. By symmetry, we can treat only the case
$\bk=0$. Around $\bk =0$, $\psi (\bk) \sim a |\bk|^2$ and
$(\partial_{\kappa^1} \omega (\bk))^2\sim b (\nu +
|\bk|^2)^{-1}(\bk^1)^2$ where $a$ and $b$ are positive constants
depending on $\nu$ and $\alpha$. Essentially, $C_{1,1} (t)$ has the
same behavior as 
\begin{equation}
\int_{\bk \in [0,1]^d} d \bk \cfrac{ (k^1)^2 e^{-a\gamma t
    |\bk|^2}}{\nu + |\bk|^2} 
=\cfrac{1}{t^{d/2 +1}} \int_{[0,\sqrt{t}]^d} d\bk \cfrac{(\bk^1)^2
  e^{-a \gamma |\bk|^2}}{\nu + t^{-1} |\bk |^2} 
\end{equation}  

Hence, we have proved the following theorem

\begin{theo}
\label{thm-correlation}
In the $(\alpha,\nu)$-harmonic case, the current-current time correlation function $C_{1,1} (t)$ decays
like
\begin{itemize}
\item $C_{1,1} (t) \sim t^{-d/2}$ in the unpinned case ($\nu =0$)
\item  $C_{1,1} (t) \sim t^{-d/2-1}$ in the pinned case ($\nu >0$)
\end{itemize}
\end{theo}

%%%%%%%%%%%%%%%%%%%%%%%%%%%%%%%%%%%%%%%%%%%%%%%%%%
%%%%%%%%%%%%%%%%%%%%%%%%%%%%%%%%%%%%%%%%%%%%%%%%%%

\section{Conductivity in the harmonic case}
\label{harm2}

%%%%%%%%%%%%%%%%%%%%%%%%%%%%%%%%%%%%%%%%%%%%%%%%%%
%%%%%%%%%%%%%%%%%%%%%%%%%%%%%%%%%%%%%%%%%%%%%%%%%%
%%%%%%%%%%%%%%%%%%%%%%%%%%%%%%%%%%%%%%%%%%%%%%%%%%

\begin{lemma}
\label{lem:inversion}
Consider the $(\alpha,\nu)$-harmonic case. For any time $t$, the following limit exists 
\begin{equation}
C_{1,1} (t)=\lim_{N \to \infty} \frac 1{N^d} \ME(\frak
J_{\be_1}(t)\frak J_{\be_1}(0)) 
\end{equation}
and
\begin{equation}
\label{eq:A13}
 \int_0^\infty dt e^{-\lambda t} C_{1,1} (t) \; dt= \lim_{N \to
   \infty}
 \left<j^a_{0,\be_1}u_{\lambda,N}\right>_{\scriptscriptstyle{N,\mathcal
     E}}  
\end{equation}
The same result holds with $\ME$ replaced by $\CE$.
\end{lemma}

\begin{proof}
We only prove this lemma in the microcanonical setting. Let us define
\begin{equation}
f_N (t)=  \frac 1{N^d} \ME (\frak J_{\be_1}(t)\frak J_{\be_1}(0))
\end{equation}
We first prove the sequence $(f_N)_N$ is uniformly bounded. By Cauchy-Schwarz and stationarity, we have
\begin{equation}
\begin{array}{lcl}
\left|f_N (t) \right| &\leq& \frac 1{N^d} \sqrt{\left<\frak
    J^2_{\be_1}(t) \right>_{\scriptscriptstyle{N,\mathcal E}}}
\sqrt{\left<\frak J^2_{\be_1}(0) \right>_{\scriptscriptstyle{N,\mathcal E}}}\\ 
&=& \frac 1{N^d} \left<\frak J^2_{\be_1}\right>_{\scriptscriptstyle{N,\mathcal E}}
\end{array} 
\end{equation}
We now use symmetry properties of the microcanonical ensemble to show
this last term is bounded above by a constant independent of $N$.  
\begin{eqnarray*}
N^{-d}<\frak J^2_{\be_1}>_{\scriptscriptstyle{N,\mathcal E}} = \sum_{\bx}
<j^a_{0,\be_1} j^a_{\bx,\bx +\be_1}>_{\scriptscriptstyle{N,\mathcal E}} \\ 
= \cfrac{\alpha^2}{4} \sum_{\bx} \sum_{i,j=1}^d \left< (\bq_{\be_1}^i
  -\bq_{0}^i )({\bq}_{\bx +\be_1}^j -\bq_{\bx}^j) (\bp_{\be_1}^i
  +\bp_0^i) (\bp_{\bx+\be_1}^j +\bp_{\bx}^j
\right>_{\scriptscriptstyle{N,\mathcal E}} 
\end{eqnarray*}
%Define 
%$$K_N (\bq)= N^d e -\cfrac{1}{2} \sum_{\bx} \bq_{\bx}\cdot (\nu I-\alpha \Delta) \bq_{\bx}$$
In the unpinned case $\nu=0$, conditionally to the positions
configuration $\bq$, 
the law of $\bp$ is $\mu_{\bq}=\mu_{\sqrt{2K_N (\bq)}}^{N^d}$ (defined
in lemma \ref{lem:sym2}).
% meaning the uniform measure on the surface 
%$$ \left\{(\bp_\bx)_{\bx \in {\mathbb Z}_N^d}; \quad \cfrac{1}{2}\sum_{\bx} \bp_\bx^2 = K_N (\bq); \quad \sum_{\bx} \bp_{\bx} =0  \right\}$$ 

By using properties (i),(ii) and (iii) of lemma \ref{lem:sym2}, one has
\begin{equation}
N^{-d}<\frak J^2_{\be_1}>_{\scriptscriptstyle{N,\mathcal E}}
=\cfrac{\alpha^2}{4} \sum_{i=1}^d \left< (\bq_{\be_1}^i -\bq_{0}^i
  )({\bq}_{2\be_1}^i -\bq_{-\be_1}^i-3\bq_{\be_1}^i +3 \bq_0^i)
  (\bp_0^i)^2\right>_{\scriptscriptstyle{N,\mathcal E}} 
\end{equation}
By Cauchy-Schwarz inequality, the modulus of this last quantity is bounded above by
\begin{equation}
\alpha [8<\cE_{0}^2>_{\scriptscriptstyle{N,\mathcal E}} +\cfrac{1}{2}
<\cE_{\be_1}^2>_{\scriptscriptstyle{N,\mathcal E}}]= \cfrac{17}{2}
<{\cE}_{0}^2>_{\scriptscriptstyle{N,\mathcal E}} 
\end{equation}
where the last equality is a consequence of the invariance by
translation of $<\cdot>_{\scriptscriptstyle{N,\mathcal E}}$.  Let
$(X_1,\ldots,X_{N^d})$ be a random vector with law
$\lambda^{N^d}_{\sqrt{N^d \mathcal E}}$, meaning the uniform measure
on the $N^d$-dimensional sphere of radius $\sqrt{N^d \mathcal E}$. The
vector of energies $({\cE}_{\bx}, \bx \in {\mathbb Z}_{N}^d)$  has the
same law as $ (X_1^2,\ldots, X_{N^d}^2)$. By lemma
\ref{lem:df-unbounded}, ${\mathbb
  E}(X_1^4)=<{\cE}_{0}^2>_{\scriptscriptstyle{N,\mathcal E}}$ is bounded above
by a constant independant of $N$. Hence there exists a positive
constant $C$ such that  
\begin{equation}
\label{eq:f}
|f_N (t)| \leq C
\end{equation}
Similarly, inequality (\ref{eq:f}) can be proved in the pinned case $\nu>0$. Let $f(t)$ be any limit point of the sequence $(f_N (t))_{N\geq 1}$ and choose a subsequence $(N_k)_{k\geq 0}$ such that $(f_{N_k})$ converges to $f$ (for the pointwise convergence  topology). By Lebesgue's theorem, we have 
\begin{equation}
\lim_{k \to \infty} \int_{0}^{\infty} e^{-\lambda t} f_{N_k} (t) dt = \int_0^\infty e^{-\lambda t} f(t) dt 
\end{equation}
But we have that
\begin{equation}
\int_{0}^{\infty} e^{-\lambda t} f_{N} (t) dt = -<j_{0,\be_1}, u_{\lambda,N}>_{\scriptscriptstyle{N,\mathcal E}} 
\end{equation}
and we have seen in section \ref{sec:cfe} this last quantity converges as $N$ goes to infinity to
\begin{equation}
\int_{0}^\infty e^{-\lambda t} f_{\infty} (t) dt
\end{equation}
where $f_{\infty}$ is given by (see (\ref{eq:dr}-\ref{eq:psi}) for the notations)
\begin{equation}
 f_{\infty} (t)=\cfrac{\mathcal E^2}{4\pi^2 d}\int_{[0,1]^d} (\partial_{\bk^1} \omega (\bk))^2 e^{-t\gamma \psi (\bk)} d\bk
\end{equation}
By injectivity of the Laplace transform, we get $f(t)=f_{\infty} (t)$. Uniqueness of limit points implies $(f_N (t))_{N\geq 1}$ converges to $f_{\infty} (t)$ for any $t$. 
It follows also we can inverse time integral and infinite volume limit
in the left hand side of (\ref{eq:A13}) and the lemma is proved. 
\end{proof}

\begin{lemma} (Resolvent equation)
\label{lem:u}
\begin{equation*}
  \label{eq:u}
   u_{\lambda,N} = \left(\lambda - L\right)^{-1} \left(- \sum_\bx
   j^a_{\bx,\bx+\be_1}\right) = \frac \alpha\gamma \sum_{\bx,\by}
 g_{\lambda,N} (\bx - \by) \bp_\bx\cdot \bq_{\by} 
\end{equation*}
where $g_{\lambda,N}(\bz)$ is the solution (such that $\sum_{\bz} g_{\lambda,N} (\bz) =0$)
of the equation 
\begin{equation}
  \label{eq:g}
  \frac{2\lambda}{\gamma} g_{\lambda,N} (\bz)- 4\Delta g_{\lambda,N}
  (\bz) =  (\delta(\bz +\be_1) -  \delta(\bz - \be_1) )
\end{equation}
for $d\ge 2$, or
\begin{equation}
\label{eq:g1d}
   \frac{2\lambda}{\gamma} g_{\lambda,N} (z) - \frac 13\Delta\left[ 4
     g_{\lambda,N} (z) +  g_{\lambda, N} (z+1) + g_{\lambda,N} (z-1)\right]
   =  (\delta(z + 1) - \delta(z - 1) )
\end{equation}
for $d=1$.
Moreover, $A u_{\lambda,N} =0$ and $L u_{\lambda,N} = \gamma S u_{\lambda,N}$.
\end{lemma}

\begin{proof}
We only give the proof for the dimension $d\geq 2$ since the proof for the
one dimensional case is similar. Let $u_{\lambda,N}=\frac \alpha\gamma
\sum_{\bx,\by} g_{\lambda,N} (\bx - \by) \bp_\bx\cdot \bq_{\by}$. The generator
$L$ is equal to the sum of the Liouville operator $A$ and of the noise
operator $\gamma S$. The action of $A$ on $u_{\lambda,N}$ is null. Indeed, we
have: 
\begin{equation}
A u_{\lambda,N} = \cfrac{\alpha}{\gamma} \sum_\bx  [ (\alpha\Delta - \nu
I)\bq_\bx] \cdot \left(\sum_\by g_{\lambda,N} (\bx -\by) \bq_\by \right)+
\cfrac{\alpha}{\gamma} \sum_{\by,\bx} g_{\lambda,N} (\bx -\by) \bp_\bx \cdot
\bp_\by  
\end{equation}
Here, and in the sequel of the proof, sums indexed by $\bx,\by,\bz$
are indexed by ${\mathbb Z}_N$ and sums indexed by $i,j,k,\ell$ are
indexed by $\{1,\ldots,d\}$. Summation by parts can be performed
(without outcoming boundary terms since we are on the torus) and we
get 
\begin{equation}
A u_{\lambda,N} = \cfrac{\alpha}{\gamma} \sum_\bx  [ (\alpha\Delta - \nu I) g_{\lambda,N}]
(\bx -\by) \bq_x \bq_\by + \cfrac{\alpha}{\gamma} \sum_{\by,\bx} g_{\lambda,N}
(\bx -\by) \bp_\bx \cdot \bp_\by  
\end{equation}
Remark now that the function $\delta (\cdot -\be_1) -\delta (\cdot
+\be_1)$ is antisymmetric. Hence $g_{\lambda,N}$, and consequently
$\Delta g_{\lambda,N}$,  is still antisymmetric. We have therefore $A
u_{\lambda,N}$ which is of the form: 
\begin{equation}
A u_{\lambda,N} = \sum_{\bx,\by} \{a_1 (\bx-\by) \bp_\bx\cdot \bp_\by
+ a_2 (\bx - \by) \bq_\bx\cdot \bq_\by\} 
\end{equation}
with $a_1,a_2$ antisymmetric. Using the antisymmetricity of $a_1$ and
$a_2$, it is easy to show that the last two sums are zero and hence $A
u_{\lambda,N} =0$.\\ 

A simple computation shows that if $\ell \in \{1,\ldots,d\}$ then
\begin{eqnarray*}
S(\bp_\bx^{\ell})&=&\cfrac{1}{2(d-1)} \sum_{\by} \sum_{i\neq j,k} (X_{\by,\by +\be_k}^{i,j})^2 (\bp_\bx ^\ell)\\
&=&\cfrac{2}{2(d-1)} \sum_{i\neq \ell,k} (X_{\bx,\bx +\be_k}^{i,\ell})^2 (\bp_\bx ^\ell)+\cfrac{2}{2(d-1)} \sum_{i\neq \ell,k} (X_{\bx-\be_k,\bx}^{i \neq\ell, k})^2 (\bp_\bx ^\ell) \\
&=&\cfrac{1}{d-1} \sum_{i\neq \ell, k} \left\{(\bp_{\bx + \be_k}^{\ell} - \bp_{\bx}^{\ell}) - (\bp_{\bx}^{\ell} - \bp_{\bx - \be_k}^{\ell})\right\}\\
&=&2 \Delta (\bp_{\bx}^\ell )
\end{eqnarray*}
Since the action of $S$ is only on the $\bp$'s, we have
\begin{eqnarray*}
\gamma S u_{\lambda,N} &=& \alpha \sum_{\bx, \by} g_{\lambda,N} (\bx
-\by) S (\bp_\bx)\cdot \bq_\by\\ 
&=&2\alpha \sum_{\bx, \by} g_{\lambda,N} (\bx - \by) (\Delta \bp
_\bx)\cdot \bq_\by\\ 
&=&2\alpha \sum_{\bx, \by} (\Delta g_{\lambda,N})(\bx -\by) \bp_\bx \cdot \bq_\by
\end{eqnarray*}
where in the last line, we performed a summation by parts. Since
$g_{\lambda,N}$ is solution of (\ref{eq:g}), we have 
\begin{equation}
{\lambda} u_{\lambda,N} - \gamma S u_{\lambda,N} =
\cfrac{\alpha}{2} \sum_\bx \bp_\bx \cdot 
(\bq_{\bx + \be_1} - \bq_{\bx - \be_1})= -\sum_{\bx} j^{a}_{\bx, \bx
  +\be_1} 
\end{equation}
\end{proof}

Let us define the Fourier transform ${\hat v}(\xi),\, \xi \in
{\mathbb Z}_N^d$, of the function $v$ on ${\mathbb Z}_N^d$ as 
\begin{equation}
\label{ft}
{\hat v}(\xi)= \sum_{\bz \in {\mathbb Z}_N^d} v(\bz)\exp(2i\pi \xi\cdot \bz/N)
\end{equation}
The inverse transform is given by
\begin{equation}\label{ift}
{ v}(\bz)= \frac 1{N^d} \sum_{\xi \in {\mathbb Z}_N^d} \hat v(\xi)
\exp(- 2i\pi \xi\cdot \bz/N)
\end{equation}
On $\mathbb Z^d$ we define similarly:
\begin{equation}
\label{ftz}
{\hat v}(\bk)= \sum_{\bz \in \mathbb Z^d} v(\bz)\exp(2i\pi \bk\cdot
\bz),  \qquad \bk \in \mathbb [0,1]^d
\end{equation}
and its inverse by
\begin{equation*}
  { v}(\bz)= \int_{[0,1]^d} \hat v(\bk)
\exp(- 2i\pi \bk\cdot \bz)
\end{equation*}

For $\lambda>0$, the function $g_{\lambda}: \Z^d \to \R$ is the solution on $\Z^d$ of the equation
\begin{eqnarray}
  \label{eq:15}
    \cfrac{2\lambda}{\gamma} g_{\lambda}(\bz) - 4 \Delta g_{\lambda}(\bz) = \delta_0(\bz +\be_1) -\delta_0 (\bz +\be_1),\quad d\ge 2\\
     \cfrac{2 \lambda}{\gamma} g_{\lambda}(z) - \frac 13 \Delta \left(4g_{\lambda}(z) +g_{\lambda}(z+1)+ g_{\lambda}(z-1)\right)  = \delta_0 (z+1) - \delta_0 (z-1) \nonumber \\
\quad d=1 \nonumber
\end{eqnarray}
%and in all cases, we have $g_{\eta,N}(\bx) =  G_{\eta,N}( \bx + \be_1) -G_{\eta,N}( \bx - \be_1)$. 

Then we have
\begin{equation}
\label{eq:gg2}
  \hat g_{\lambda} (\bk) = \frac{-2i\pi \sin (2\pi \bk^1)}{\cfrac{2\lambda}{\gamma} + 16 \sum_{j=1}^d \sin^2(\pi \bk^j)}, \qquad \text{if $d\ge 2$}
\end{equation}
and 
\begin{equation}
\label{eq:gg1}
  \hat g_{\lambda} (k) = \frac{-2i\pi \sin(2\pi k)}
  {\cfrac{2 \lambda}{\gamma} + \cfrac{8}{3} \sin^2(\pi k)\left(1 +2\cos^2(\pi k)\right)}, 
    \qquad \text{if $d = 1$}
\end{equation}

Since $g_{\lambda,N}$ is the solution of the same equation as $g_\lambda$ but on $\To_N^d$, we have the following formula for ${\hat g}_{\lambda,N}$:
\begin{equation}
{\hat g}_{\lambda,N} (\xi)={\hat g}_{\lambda} (\xi/N)
\end{equation}

%Since
%\begin{equation*}
 %  \hat g_{\eta,N} (\xi) = -2 i \sin(2\pi \xi^{1} /N) \hat G_{\eta,N} (\xi) 
%\end{equation*}
The following bound follows easily from Parseval relation:
\begin{equation}
  \label{eq:16}
  \sum_{\bx \in {\mathbb Z}_N^d} ( g_{\lambda,N} (\bx))^2 \leq \frac{\gamma^2}{\lambda^2}
\end{equation}

Similarly, the function $\Gamma_N$ defined in (\ref{eq:gamma}) has Fourier transform given by
\begin{equation}
\hat \Gamma_N (\xi) = \hat\Gamma (\xi/N)
\end{equation}
where
\begin{equation}
  \hat \Gamma (\bk) = \frac {-2i \sin (2\pi \bk^1)}{\nu + 4 \alpha  \sum_{j=1}^d \sin^2(\pi \bk^j)}
\end{equation}

%and it follows the well known estimates
%\begin{equation}
 % \label{eq:satd}
 % \begin{array}{lll}
%&\Gamma_N (0) \quad \text{is of order $N$ if $d=1, \nu=0$.}\\
%&\Gamma_N (0) \quad \text{is of order $\log N$ if $d=2, \nu=0$.}\\
%&\Gamma_N (0) \quad \text{is of order $1$ if $d\geq 3$ or $\nu>0$.}\\
%\end{array}
%\end{equation}

%As $N\to \infty$ we have $G_{\eta,N} \to G_\eta$ (resp. $g_{\eta,N} \to g_\eta$) solution of the same equations (\ref{eq:15}) (resp. (\ref{eq:g},\ref{eq:g1d})) but on $\mathbb Z^d$. The convergence is understood pointwise and in ${\mathbb L}^2$. 

%We can compute then, for $d=1$, 
%\begin{equation}
%  \label{eq:GG01}
%   G_\eta(0) +  G_\eta(1) = \gamma \int_0^1 \frac{\cos^2(\pi k)}{\eta +
 %    \gamma/3 \sin^2(\pi k) \left(1+2\cos^2(\pi k)\right)} dk
%\end{equation}

Let us denote by $z^*$ the conjugate of the complex number $z$ and observe that the function $\bk \in {[0,1]^d} \to {\hat g}_\lambda (\bk) \left[{\hat \Gamma} (\bk)\right]^* \in {\R}^+$ is continuous.  Hence we have the following convergence of Riemann sums
\begin{eqnarray}
\label{eq:conv-RS}
\sum_{\by \in \To_N^d} g_{\lambda,N} (\by) \Gamma_N (\by) = \cfrac{1}{N^d} \sum_{\xi \in \To_N^d} {\hat g}_{\lambda,N}(\xi) [{\hat \Gamma}_N (\xi)]^*\\
{\xrightarrow[N\to \infty]{}}\int_{[0,1]^d} d\bk {\hat g}_{\lambda}(\bk) [{\hat \Gamma} (\bk)]^*= \sum_{\by \in \Z^d} g_{\lambda} (\by) \Gamma (\by) \nonumber
\end{eqnarray}

The limits as $\lambda \to 0$ of the above expressions give the values for the conductivity (up to a multiplicative constant) when this is finite. If $\nu = 0$ it diverges if $d= 1$ or $2$. 

%In these last cases, it is of some interest to obtain the behavior of the conductivity with the size $N$ of the system. A method consists to study $\kappa_N^{1,1}$ defined by inverting the temporal and spatial limits in (\ref{eq:gk0}):
%\begin{equation}
%\kappa_N^{1,1}=\lim_{t\to\infty} \frac 1{2 e^2 t} \frac 1{N^{d}}\mathbb E
%\left(\left[\sum_\bx J_{\bx,\bx+\be_1}(t)\right]^2 \right) 
%\end{equation}
%Following the computations developped in the present paper, it is not difficult to show that $\kappa_N^{1,1}$ is given by 
%\begin{equation}
%\kappa_N^{1,1}= \frac {\alpha^2 e^2}{4 d \gamma} \sum_\bz g_N (\bz) \left(\Gamma_N (0,\bz+ \be_1)) - \Gamma_N (0,\bz - \be_1)\right)
%\end{equation}
%where $g_{N}=\lim_{\lambda \to 0} g_{\lambda,N}$. One obtains then a divergence like $N$ in $1$ dimensional unpinned case and like $\log N$ in the $2$ dimensional case. In other cases, $\kappa_N^{1,1} \to \kappa^{1,1}$. In the diverging cases, it is not clear the behavior obtained coincides with the behavior of the conductivity for the finite system of length $N$ with thermic baths at the boundary.  

%%%%%%%%%%%%%%%%%%%%%%%%%%%%%%%%%%%%%%%%%
%%%%%%%%%%%%%%%%%%%%%%%%%%%%%%%%%%%%%%%%%%
%%%%%%%    Anharmonic case     %%%%%%%%%%%%%%%%%%%%%%%%%%%%
%%%%%%%%%%%%%%%%%%%%%%%%%%%%%%%%%%%%%%%%%%%%
%%%%%%%%%%%%%%%%%%%%%%%%%%%%%%%%%%%%%%%%%%%%%

\section{Anharmonic case: bounds on the thermal conductivity}
\label{sec:anharm-case:-bounds}

We consider in this section the general anharmonic case and we prove
theorem \ref{th-anharm}.
Recall \eqref{eq:5c}, then all we need to estimate is
\begin{equation}
\label{eqtobebond}
   (2 T^2 N^{d+1})^{-1} \CE\left(\left[\sum_\bx \int_0^N
        j^a_{\bx,\bx+\be_1}(s) ds\right]^2 \right)
\end{equation}

Let us define $\sum_\bx j^a_{\bx,\bx+\be_1} = {\frak J}_{\be_1}$, then we have
the general bound (\cite{S}, lemma 3.9)
\begin{equation}
\label{eq:755}
\begin{array}{lcl}
    \CE \left(\left[ \int_0^N {\frak J}_{\be_1}(s) ds\right]^2 \right)
    &\le& 10 N \left< {\frak J}_{\be_1}, (N^{-1} - L)^{-1} {\frak J}_{\be_1} \right>_{\scriptscriptstyle{N,T}}\\
    &\le& 10 N \left< {\frak J}_{\be_1}, (N^{-1} - \gamma S)^{-1} {\frak J}_{\be_1} \right>_{\scriptscriptstyle{N,T}}\\
 %   &=&10 t \gamma^{-1} \left< {\frak J}_{\be_1}, (\, (t\gamma)^{-1} - S)^{-1} {\frak J}_{\be_1} \right>_{\scriptscriptstyle{N,T}}
\end{array}
\end{equation}

Recall that $S(\bp_\bx)=2\Delta(\bp_\bx)$ if $d\geq 2$ and $S(p_x)=\cfrac{1}{6} \Delta(4p_x +p_{x +1} +p_{x -1})$ if $d=1$. 
\begin{equation}
(N^{-1}- \gamma S)^{-1} {\frak J}_{\be_1}= \sum_{j=1}^d \sum_{\by} G_N (\bx -\by)\bp_x^j V_{j}' (\bq^j_{\by +\be_1} -\bq_{\by}^j) 
\end{equation}
where $G_N (\bz)$ is the solution of the resolvent equation
\begin{equation}
\label{eq:G}
\begin{cases}
N^{-1} G_N (\bz) -2\gamma (\Delta G_N)(\bz)=-\cfrac{1}{2}\left[\delta_0 (\bz) +\delta_{\be_1} (\bz)\right], \quad d\geq 2\\
\begin{array}{ll}
\phantom{=}&N^{-1} G_{N} (\bz) -\cfrac{\gamma}{6}\left[4(\Delta G_{N})(\bz)+(\Delta G_{N})(\bz +1) +(\Delta G_{N})(\bz -1)\right]\\
=&-\cfrac{1}{2}\left[\delta_0 (\bz) +\delta_{1} (\bz)\right], \quad d=1
\end{array}
\end{cases}
\end{equation}

The left hand side of (\ref{eq:755}) is equal to
\begin{equation}
\label{eq:775}
  \begin{split}
    -5 T N^{d+1} \sum_{j=i}^d \sum_{\bx} \left( G_{N}(\bx) + G_{N}(\bx + \be_1) \right) \left< V'_j (q^j_{\bx + \be_1} - q^j_\bx) V'_j(  q^j_{ \be_1} - q^j_0) \right>_{\scriptscriptstyle{N,T}} 
  \end{split}
\end{equation}

\begin{itemize}
\item \textbf{Pinned case}\\

In the pinned case, the correlations $ \left< V'_j(q^j_{\bx + \be_1} - q^j_\bx)
      V'_j( q^j_{ \be_1} - q^j_0) \right>_{\scriptscriptstyle{N,T}}$ 
decay exponentially in $\bx$
\begin{equation}
\label{eq:deco2}
\left|\left< V'_j(q^j_{\bx + \be_1} - q^j_\bx) V'_j( q^j_{ \be_1} - q^j_0) \right>_{\scriptscriptstyle{N,T}}\right | \le Ce^{-c|\bx|}
\end{equation}

It follows that the previous expression is
bounded by 
\begin{equation*}
  C T^2 t N^d \sum_\bx |G_{N}(\bx) +  G_{N}(\bx+ \be_1)| e^{-c|\bx|} 
\end{equation*}
Since $G_{N}$ is bounded in $d\ge 3$, it follows that (\ref{eqtobebond}) is uniformly bounded in $N$. In low dimensions, our estimate are to rough and we obtain only diverging upper-bounds. Nevertheless, if $V_j (r)= \alpha_j r^2$ are quadratics and $W_j$ are general but strictly positive then 
\begin{equation}
\begin{split}
 \left< V'_j(q^j_{\bx + \be_1} - q^j_\bx) V'_j( q^j_{ \be_1} - q^j_0) \right>_{\scriptscriptstyle{N,T}}\\
 =\alpha_j \left\{ 2<\bq_\bx^j \bq_0^j>_{\scriptscriptstyle{N,T}}- <\bq_{\bx-\be_1}^j \bq_0^j>_{\scriptscriptstyle{N,T}}-<\bq_{\bx+\be_1}^j \bq_0^j>_{\scriptscriptstyle{N,T}} \right\}
\end{split}
\end{equation}
As a function of $\bx$, this quantity is a Laplacian in the first direction and by integration by parts, the left-hand side of (\ref{eqtobebond}) is upper bounded by
\begin{equation}
C \sum_\bx \left |(\Delta G_{N}) (\bx) + (\Delta G_{N}) (\bx + \be_1)\right|e^{-c|\bx|}
\end{equation}
By lemma \ref{lem:Green}, this quantity is uniformly bounded in $N$.
\newline

\item \textbf{Unpinned case}\\

In the unpinned case, we assume that  $0 < c \le V_j''(q) \le C < +\infty$. We have
(cf. \cite{dd}, theorem 6.2, that can be proved in finite volume uniformly)
\begin{equation}
\label{eq:deco1}
   \left|\left< V'_j(q^j_{\bx + \be_1} - q^j_\bx)
      V'_j(q^j_{ \be_1} - q^j_0) \right>_{\scriptscriptstyle{N,T}}\right| \le C |\bx|^{-d}
\end{equation}
In the one dimensional case, the random variables $r_x = q_{x+1}-q_x$ are i.i.d. and $<V'(r_x)>_{\scriptscriptstyle{N,T}} =0$. Only the term corresponding to $x=0$ remains in the sum of (\ref{eq:775}). By lemma \ref{lem:Green}, we get the upper bound
\begin{equation}
\left(G_{N} (0) +G_{N}(1)\right) \left< V' (r_0^2)\right>_{\scriptscriptstyle{N,T}} \leq C {\sqrt N}
\end{equation}

For the unpinned two dimensional case, we obtain the upper bound
\begin{equation}
\begin{split}
C \sum_{\bx \in \To_N^2} |G_{N}(\bx) +  G_{N}(\bx+ \be_1)| |\bx|^{-d}\\
\leq C \log N \sum_{\bx \in \To_N^2} |x|^{-2}\\
\sim C(\log N)^2
\end{split}
\end{equation}

For the case $d\geq 3$, we use the first point of lemma \ref{lem:Green}, (\ref{eq:deco1}) and the fact that
\begin{equation}
\sum_{\bx \in {\To}_N^d} |\bx|^{-d} \sim \log N
\end{equation}

\end{itemize}

\begin{lemma}
\label{lem:Green}
Let $G_N$ be the solution of the discrete equation (\ref{eq:G}). There exists a constant $C>0$ independent of $N$ such that
\begin{itemize}
\item $G_N (\bx) \leq C( |\bx|^{d-2}+ N^{-1/2}), \quad d\geq 3$
\item $G_N (\bx) \leq C \log N, \quad d = 2$
\item $G_N (\bx) \leq  C\sqrt{N}, \quad d =1$
\item $|G_N (\bx +\be_1)+G_N (\bx -\be_1) -2 G_N (\bx)| \leq C, \quad d\geq 1$

\end{itemize}
\end{lemma}

\begin{proof}

In the proof, $C$ is a constant independent of $N$ but which can change from line to line. We first treat the case $d\geq 3$. We use Fourier's transform representation of $G_N$:
\begin{equation}
G_N (\bx)= -\cfrac{1}{2 N^d} \sum_{\bk \in \To_N^d} (1+e^{2i\pi \bk^1/N}) \cfrac{e^{-2i\pi \bk\cdot \bx/N}}{\theta_N (\bk/N)}
\end{equation}
where $\theta_N (\bu)=N^{-1}+8\gamma \sum_{j=1}^d \sin^{2} (\pi \bu^j)$. $G_N$ can also be written in the following form
\begin{equation}
G_N (\bx)=-\cfrac{1}{2} \left[ F_N (\bx) +F_N (\bx -\be_1)\right]
\end{equation}
where
\begin{equation}
F_N (\bx)=\cfrac{1}{N^d} \sum_{\bk \in \To_N^d}\cfrac{e^{-2i\pi \bk\cdot \bx/N}}{\theta_N (\bk/N)}
\end{equation}

Let us introduce the continuous Fourier's transform representation of the Green function $F_\infty$ on $\Z^d$ given by:
\begin{equation}
F_{\infty} (x) =\int_{[0,1]^d} \cfrac {\exp(2i\pi \bx \cdot \bu)}{\theta (\bu)}d\bu
\end{equation}
where $\theta (\bu) =8 \gamma  \sum_{j=1}^d \sin^2 (\pi \bu^j)$. Remark that $F_{\infty}$ is well defined because $d\geq 3$. We have to prove there exists a constant $C>0$ independent of $N$ such that
\begin{equation}
F_N (\bx) \leq C( |\bx|^{d-2}+ N^{-1/2})
\end{equation}

Observe that by symmetries of $F_N$, we can restrict our study to the case $\bx \in [0,N/2]^d$.

We want to show that $F_N (\bx)$ is well approximated by $F_{\infty} (\bx)$. We have
\begin{equation}
F_N (\bx) - F_{\infty} (\bx) =F_N (\bx) -F_{\infty}^{N} (\bx) +F_{\infty}^N (\bx) -F_{\infty} (\bx)
\end{equation}
where
\begin{equation}
F_{\infty}^N (\bx) =\int_{[0,1]^d} \cfrac {\exp(2i\pi \bx \cdot \bu)}{\theta_N (\bu)}d\bu
\end{equation}

For each $\bk \in \To_N^d$, we introduce the hypercube $Q_\bk=\prod_{j=1}^d [\bk^j/N, (\bk^j+1)/N)$ and we divide $[0,1]^d$ following the partition $\cup_{\bk \in \To_N^d} Q_\bk$. By using this partition, we get
\begin{equation}
\label{eq:green82}
\begin{array}{lcl}
F_N (\bx)-F_\infty^{N} (\bx) &=& \sum_{\bk \in \To_N^d} \int_{Q_\bk} d\bu\cfrac{e^{2i\pi \bk \cdot \bx/N}-e^{2i\pi \bu \cdot \bx}}{\theta_{N} (\bk /N)}\\
&+&\int_{Q_{\bk}}{ d\bu e^{2i\pi \bu\cdot \bx}\left( \cfrac{1}{\theta_N (\bk /N)} -\cfrac{1}{\theta_N (\bu)}\right)}
\end{array}
\end{equation}

Remark that
\begin{equation}
\int_{Q_\bk} d\bu e^{2i\pi \bu \cdot \bx}=\cfrac{e^{2i\pi \bk \cdot \bx /N}}{N^d} \varphi (\bx/N)
\end{equation}
where 
\begin{equation}
\varphi (\bu)=\prod_{j=1}^{d} e^{2i\pi \bu^j} \prod_{j=1}^{d}\cfrac{\sin(\pi \bu^j)}{(\pi \bu^j)}
\end{equation}
It follows that the first term on the right hand side of (\ref{eq:green82}) is equal to 
\begin{equation}
(1-\varphi (\bx/N)) F_N (\bx)
\end{equation}
so that
\begin{equation}
\label{eq:green86}
F_N (\bx) = \cfrac{F_{\infty}^N (\bx)}{\varphi (\bx /N)} + \cfrac{1}{\varphi (\bx /N)} \sum_{\bk \in \To_N^d} \int_{Q_{\bk}} d\bu e^{2i\pi \bu\cdot \bx}\left( \cfrac{1}{\theta_N (\bk /N)} -\cfrac{1}{\theta_N (\bu)}\right)
\end{equation}
The next step consists to show that the second term on the right hand side of (\ref{eq:green86}) is small. In the sequel, $C$ is a positive constant independent of $N$ but which can change from line to line. For each $\bu \in Q_{\bk}$, we have
\begin{equation}
\sin^2 (\pi \bu^j) - \sin^2 (\pi \bk^j/N)= \pi \sin (2\pi c_j) (\bu^{j} -\bk^j/N)
\end{equation}
for some $c_j \in [\bk^j/N, (\bk^{j}+1)/N)$. Consequently, we have
\begin{equation}
|\sin^2 (\pi \bu^j) - \sin^2 (\pi \bk^j/N)| \leq \cfrac{C}{N} |\sin(\pi \bk^j /N)|
\end{equation}
Moreover, there exists a positive constant $C$ such that
\begin{equation}
\forall \bk \in \To_N^d, \forall \bu \in Q_\bk, \quad \theta_N (\bu) \geq C\theta_N (\bk/N)
\end{equation}
It follows that the modulus of the second term on the right hand side of (\ref{eq:green86}) is bounded by 
\begin{equation}
\cfrac{C}{|\varphi (\bx/N)|}\sum_{j=1}^d \cfrac{1}{N^d} \sum_{\bk \in \To_N^d} \cfrac{N^{-1}  |\sin (\pi \bk^j/N)|} {\theta_N (\bk/N)^2}
\end{equation}
Since the modulus of the function $\varphi (\bu)$ is bounded below by a positive constant on $[0,1/2]^d$,this last term is of the same order as 
\begin{equation}
N^{-1} \sum_{j=1}^d \int_{[0,1]^d} \cfrac{|\sin(\pi \bu^j)|}{\theta_N (\bu)^2}d\bu
\end{equation}
Elementary standard analysis shows that this term is of the same order as
\begin{equation}
N^{-1} \int_0^1 \cfrac{r^d}{(N^{-1} +r^2)^2} dr
\end{equation}
For $d\geq 4$, this term is clearly of order $N^{-1}$. For $d=3$, the change of variables $r=N^{-1/2}v$ gives an integral of order $N^{-1}\log N$. In conclusion, we proved
\begin{equation}
F_N (\bx)= \cfrac{F_{\infty}^N (\bx)}{\varphi (\bx /N)} + O \left(\cfrac{\log N}{N}\right)
\end{equation}
Moreover, it is not difficult to show that
\begin{equation}
|F_{\infty} (\bx) -F_\infty^N (\bx)| \leq C N^{-1/2}
\end{equation}
Since we have (cf. \cite{M}, theorem 4.5) 
\begin{equation}
F_{\infty} (\bx) \leq C |\bx|^{2-d}
\end{equation}
we obtained the first point of the lemma.

For the $1$ and $2$-dimensional estimates, we have that $|G_N (x)| \leq G_N (0)$ and by standard analysis, there exists a constant $C>0$ independent of $N$ such that 
\begin{equation}
G_N (0) \leq C \int_{[0,1/2]^d} d\bk \cfrac{1}{N^{-1} + \sum_{j=1}^d \sin^{2} (\pi \bk)}
\end{equation}
By using the inequality $\sin^{2} (\pi u) \geq 4u^2$, one gets $G_N (0)$ is of same order as
\begin{equation}
\int_{[0,1/2]^d} d\bk \cfrac{1}{N^{-1} + |\bk|^2}
\end{equation}
This last quantity is of order $\sqrt{N}$ if $d=1$ and $\log N$ if $d=2$.

Let us now prove the final statement. Assume $d\geq 2$ (the case $d=1$ can be proved in a similar way). We have
\begin{equation}
\begin{array}{l}
\left|G_N (\bx+\be_1)+G_N (\bx -\be_1) -2G_N (\bx)\right|\\
=\left|\cfrac{2}{N^d} \sum_{\bk \in \To_N^d} (1+e^{2i\pi \bk^1/N})\sin^2 (\pi \bk_1 /N) \cfrac{e^{-2i\pi \bk\cdot \bx/N}}{\theta_N (\bk/N)}\right| \\
\leq \cfrac{4}{N^d} \sum_{\bk \in \To_N^d} \cfrac{\sin^2(\pi \bk_1 /N)}{\theta_N (\bk/N)}\\
\leq ({2\gamma})^{-1}
\end{array}
\end{equation}

\end{proof}

%%%%%%%%%%%%%%%%%%%%%%%%%%%%%%%%%%%%%%%%%%%%%%%%%%%%%%%%%%%%%%%%%%%%%%%%%%%%
%%%%%%%%%%%%%%%%%%%%%%%%%%%%%%%%%%%%%%%%%%%%%%%%%%%%%%%%%%%%%%%%%%%%%%%%%%%%
%%%%%%%%%%%%%%%%%   EQUIVALENCE OF ENSEMBLES          %%%%%%%%%%%%%%%%%%%%%%
%%%%%%%%%%%%%%%%%%%%%%%%%%%%%%%%%%%%%%%%%%%%%%%%%%%%%%%%%%%%%%%%%%%%%%%%%%%%
%%%%%%%%%%%%%%%%%%%%%%%%%%%%%%%%%%%%%%%%%%%%%%%%%%%%%%%%%%%%%%%%%%%%%%%%%%%%
%%%%%%%%%%%%%%%%%%%%%%%%%%%%%%%%%%%%%%%%%%%%%%%%%%%%%%%%%%%%%%%%%%%%%%%%%%%%

\section{Appendix: Equivalence of ensembles}
\label{sec:equivalence}

In this part, we establish a result of equivalence of ensembles for
the microcanonical measure $<\cdot>_{\scriptscriptstyle{N,\mathcal E}}$ since it does not seem to
appear in the literature. The decomposition in normal modes permits to
obtain easily the results we need from the classical equivalence of
ensemble for the uniform measure on the sphere. This last result
proved in \cite{df} says that the expectation of a local function in
the microcanonical ensemble (the uniform measure on the sphere of
radius $\sqrt{k}$ in this context) is equal to the expectation of the
same function in the canonical ensemble (the standard gaussian measure
on ${\mathbb  R}^{\infty}$) with an error of order $k^{-1}$. In fact,
the equivalence of ensembles of Diaconis and Freedman is expressed in
terms of a very precise estimate of variation distance between the
microcanonical ensemble and the canonical ensemble. In this paper, we
need to consider equivalence of ensembles for unbounded functions and
to be self-contained we prove in the following lemma a slight
modification of estimates of \cite{df}. 

\begin{lemma}
\label{lem:df-unbounded}
Let $\lambda_{r n^{1/2}}^n$ be the uniform measure on the sphere 
$$S^{n}_{r n^{1/2}} = \left\{ (\bx_1,\ldots,\bx_n) \in {\mathbb R}^n; \sum_{\ell=1}^n x_\ell^2 =n r^2\right\}$$ of radius $r$ and dimension $n-1$ and $\lambda_{r}^{\infty}$ the Gaussian product measure with mean $0$ and variance $r^2$. Let $\theta >0$ and $\phi$ a function on ${\mathbb R}^k$ such that
\begin{equation}
|\phi (x_1,\ldots,x_k)| \leq C \left(\sum_{\ell=1}^k x_\ell^2\right)^{\theta},\quad C>0
\end{equation}  
There exists a constant $C'$ (depending on $C,\theta,k,r$) such that
\begin{equation}
\label{eq:df-unbounded}
\limsup_{n\rightarrow \infty} \quad n \left|\lambda_{r n^{1/2}}^n (\phi) - \lambda_r^{\infty}(\phi)\right|\leq C'
\end{equation}
\end{lemma} 

\begin{proof}
This lemma is proved in \cite{df} for $\phi$ positive bounded by
$1$. Without loss of generality, we can assume $r=1$ and we simplify
the notations by denoting $\lambda_{r n^{1/2}}^n$ with $\lambda^n$ and
$\lambda_r^{\infty}$ with $\lambda^{\infty}$. The law of
$(x_1+\ldots+x_k)^2$ under $\lambda^n$ is $n$ times a $\beta[k/2,
(n-k)/2]$ distribution and has density (cf \cite{df}) 
\begin{equation}
f(u)={\bf 1}_{\{0 \leq u \leq n\}}\cdot \cfrac{1}{n} \cfrac{\Gamma
  (n/2)}{\Gamma (k/2) \Gamma
  [(n-k)/2]}\left(\cfrac{u}{n}\right)^{(k/2)-1}
\left(1-\cfrac{u}{n}\right)^{((n-k)/2)-1} 
\end{equation}   
On the other hand, the law of $(x_1+\ldots+x_k)^2$ under $\lambda^\infty$ is $\chi_k^2$ with density (cf \cite{df})
\begin{equation}
g(u)=\cfrac{1}{2^{k/2} \Gamma (k/2)}e^{-u/2} u^{(k/2) -1}
\end{equation}
With these notations, we have
\begin{equation}
\left|\lambda^n (\phi) - \lambda^{\infty}(\phi)\right|\leq C\int_{0}^{\infty}u^{\theta} |f(u) - g(u)| du
\end{equation}
The RHS of the inequality above is equal to 
\begin{equation}
\label{eq:2006}
2C\int_0^\infty u^\theta \left(\cfrac{f(u)}{g(u)}-1\right)^{+} g(u)du + C\int_{0}^{\infty} u^{\theta} (g(u) - f(u))du
\end{equation}
In \cite{df}, it is proved $2\left(\cfrac{f(u)}{g(u)}-1\right)^+ \leq 2(k+3)/(n-k-3)$ as soon as $k \in \{1,\ldots,n-4\}$. The second term of (\ref{eq:2006}) can be computed explicitely and is equal to
\begin{equation}
\cfrac{\Gamma \left((2 \theta + k)/2\right)}{\Gamma (k/2)}\left[2^{\theta} - \cfrac{n^\theta \Gamma (n/2)}{\Gamma (\theta + n/2)}\right]
\end{equation}
A Taylor expansion shows that this term is bounded by $C'/n$ for $n$ large enough.
\end{proof}

We recall here the following well known properties of the uniform measure on
the sphere.   

\begin{lemma}\text{(Symmetry properties of the uniform measure on the
    sphere)}\\ 
\label{lem:sym}
Let $\lambda_r^k$ be the uniform measure on the sphere 
$$
S^{k}_r = \left\{ (\bx_1,\ldots,\bx_k) \in ({\mathbb R}^d)^k;
  \sum_{\ell=1}^k \bx_\ell^2 =r^2\right\}$$ of radius $r$ and
dimension $dk-1$.\\ 
i) $\lambda_r^k$ is invariant by any permutation of coordinates.\\
ii) Conditionaly to $\{\bx_1, \ldots,\bx_k\}\backslash \{\bx_i\}$, the
law of $\bx_i$ has an even density w.r.t. the Lebesgue measure on
${\mathbb R}^d$. 
\end{lemma}

In the same spirit, we have the following lemma.
\begin{lemma}
\label{lem:sym2}
Let $\mu_r^k$ be the uniform measure on the surface defined by
$$
M_r^k = \left\{ (\bx_1,\ldots,\bx_k) \in ({\mathbb R}^d)^k;\quad
  \sum_{\ell=1}^{k} \bx_\ell^2 = r^2 ;\quad \sum_{\ell=1}^{k} \bx_\ell
  = 0\right\} 
$$
We have the following properties:\\
i) $\mu_r^k$ is invariant by any permutation of the coordinates.\\ 
ii) If $i\neq j \in \{1,\ldots,d\}$ then for every $h,\ell \in
\{1,\ldots,k\}$ (distincts or not), $\mu_r^k (\bx_h^i \bx_\ell^j)
=0$.\\ 
iii) If $h \neq \ell \in \{1,\ldots,k\}$ and $i \in \{1,\ldots,d\}$, 
\begin{equation}
\mu_r^k (\bx_h^i \bx_\ell^i)= - \cfrac{r^2}{dk(k-1)}= -\cfrac{\mu_r^k (\bx_h^2)}{k-1}= -\cfrac{\mu_r^k (\bx_\ell^2)}{k-1}
\end{equation}
\end{lemma}
  
\begin{lemma} (\text{Equivalence of ensembles})\\
\label{lem:equiv-ens}
Consider the $(\alpha,\nu)$-harmonic case. There exists a positive
constant $C=C(d,\mathcal E)$ such that:\\ 
i) If $i \neq j$, $\left|\left< \left(\bp_0^{j}
      \bp_{\be_1}^i\right)^2\right>_{\scriptscriptstyle{N,\mathcal
      E}}-\cfrac{\mathcal E^2}{d^2}\right| \leq \cfrac{C}{N^d}$\\ 
ii) If $i \neq j$, $ \left|\left<  (\bp_0^i \bp_{\be_1}^i \bp_0^j \bp_{\be_1}^j)  \right>_{\scriptscriptstyle{N,\mathcal E}}\right| \leq \cfrac{C}{N^d}$\\
iii) For any $i$ and any $\by \in {\mathbb Z}_N^d$, we have
$$
\left|\left<\bq_{\by}^j (\bq_{-\be_1}^j -\bq_{\be_1}^j)
    (\bp_{0}^j)^2 \right>_{\scriptscriptstyle{N,\mathcal E}} -
  \left(\frac {\mathcal E}d \right)^2 \Gamma_N (\by) \right| \le \frac
{C {\log N} }{N^d}
$$ 
\end{lemma}

\begin{proof}
Let us treat only the unpinned case $\nu=0$. The pinned case is similar. We take the Fourier transform of the positions and of the momentums
(defined by (\ref{ft}))  and we define 
\begin{equation}
\label{eq:fourier-q-p}
\begin{split}
  \tilde \bq(\xi) = (1-\delta (\xi))\omega(\xi) \hat \bq(\xi), \qquad
  \tilde \bp (\xi) = N^{-d/2} (1-\delta (\xi)){\hat \bp} (\xi), \qquad \xi \in
  \mathbb Z_d^N 
\end{split} 
\end{equation}
where $\omega(\xi) = 2N^{-d/2} \sqrt{\alpha \sum_{k=1}^d
  \sin^2(\pi\xi^k /  N)}$ 
is the normalized dispersion relation. The factor $1-\delta$ in the definition
above is due to the condition $\sum_{\bx} \bp_{\bx}=\sum_{\bx}
\bq_{\bx} =0$ assumed in the microcanonical state. Then the energy can
be written as 
\begin{eqnarray*}
  \mathcal H_N &=& \frac 12 \sum_{\xi \neq 0} \left\{|\tilde \bp (\xi)|^2 + |\tilde \bq(\xi)|^2\right\}\\ 
&=& \frac 12 \sum_{\xi \neq 0} \left\{ {\mathfrak Re}^2 (\tilde \bp (\xi)) + {\mathfrak Im}^2 (\tilde \bp (\xi)) + {\mathfrak Re}^2 (\tilde \bq (\xi)) + {\mathfrak Im}^2 (\tilde \bq (\xi)\right\}
\end{eqnarray*}

Since $\bp_\bx, \bq_\bx$ are real, ${\mathfrak Re} (\tilde \bp), {\mathfrak Re} (\tilde \bq)$ are even and  ${\mathfrak Im} (\tilde \bp), {\mathfrak Im} (\tilde \bq)$ are odd:
\begin{equation}
\label{eq:reim}
\begin{split}
{\mathfrak Re} (\tilde \bp)(\xi) = {\mathfrak Re} (\tilde \bp) (-\xi),\qquad {\mathfrak Re} (\tilde \bq)(\xi) = {\mathfrak Re} (\tilde \bq) (-\xi)\\
{\mathfrak Im} (\tilde \bp)(\xi) = - {\mathfrak Im} (\tilde \bp) (-\xi),\qquad {\mathfrak Im} (\tilde \bq)(\xi) = - {\mathfrak Im} (\tilde \bq) (-\xi)\\
\end{split}
\end{equation}

On ${\mathbb Z}_N^d \backslash \{0\}$, we define the relation $\xi \sim \xi'$ if and only if $\xi = -\xi'$. Let ${\mathbb U}_N^d$ be a class of representants for $\sim$ (${\mathbb U}_N^d$ is of cardinal $(N^d -1)/2$). With these notations and by using (\ref{eq:reim}), we have
\begin{equation}
\mathcal H_N = \sum_{\xi \in {\mathbb U}_N^d} \left\{ {\mathfrak Re}^2 (\tilde \bp (\xi)) + {\mathfrak Im}^2 (\tilde \bp (\xi)) + {\mathfrak Re}^2 (\tilde \bq (\xi)) + {\mathfrak Im}^2 (\tilde \bq (\xi)\right\}
\end{equation}

It follows that in the microcanonical state, the random
variables 
$$
(({\mathfrak Re} {\tilde \bp})(\xi), ({\mathfrak Im}
{\tilde \bp})(\xi), ({\mathfrak Re} {\tilde \bq})(\xi), {\mathfrak Im}
{\tilde \bq})(\xi) )_{\xi \in {\mathbb U}_N^d}
$$ 
are distributed
according to the uniform measure on the sphere of radius $\sqrt{N^d
  \mathcal E}$ (which is not true without the restriction on the set ${\mathbb
  U}_N^d$). The classical results of equivalence of ensembles for the
uniform measure on the sphere (\cite{df}) can be applied for these
random variables.\\ 

i) By using inverse Fourier transform and (\ref{eq:reim}), we have
\begin{equation}
\left<  \left(\bp_0^{j} \bp_{\be_1}^i\right)^2\right>_{\scriptscriptstyle{N,\mathcal E}} = \cfrac{1}{N^{2d}} \sum_{\xi,\xi',\eta,\eta' \neq 0} \left<{\tilde \bp}^j (\xi){\tilde \bp}^j (\xi'){\tilde \bp}^i (\eta){\tilde \bp}^i (\eta')\right>_{\scriptscriptstyle{N,\mathcal E}} e^{-\cfrac{2i\pi \be_{1}\cdot (\eta + \eta')}{N}}
\end{equation}
It is easy to check by using (ii) of lemma \ref{lem:sym} that the only
terms in this sum which are nonzero are only for $\xi ' = -\xi$ and
$\eta= -\eta'$. One gets hence 

\begin{equation}
\left<  \left(\bp_0^{j} \bp_{\be_1}^i\right)^2 \right>_{\scriptscriptstyle{N,\mathcal E}} =
\cfrac{1}{N^{2d}} \sum_{\xi,\eta \neq 0} \left<\left|{\tilde \bp}^j
    (\xi)\right|^2 \left|{\tilde \bp}^i (\eta)\right|^2\right>_{\scriptscriptstyle{N,\mathcal E}}  
\end{equation}
Classical equivalence of ensembles estimates of \cite{df} show that
this last sum is equal to $(\mathcal E/d)^2 + O(N^{-d})$. 

ii) Similary, one has
\begin{equation}
  \begin{split}
    &\left< (\bp_0^i \bp_{\be_1}^i \bp_0^j \bp_{\be_1}^j) \right>_{\scriptscriptstyle{N,\mathcal E}}
    \\
    & =\cfrac{1}{N^{2d}}\sum_{\xi,xi',\eta,\eta' \neq 0} \left<{\tilde
        \bp}^i (\xi) {\tilde \bp}^i (\xi') {\tilde \bp}^j (\eta)
      {\tilde \bp}^j (\eta') \right>_{\scriptscriptstyle{N,\mathcal E}} \exp\left(-\cfrac{2i\pi
        \be_1}{N} \cdot (\xi' + \eta')\right)
  \end{split}
\end{equation}
It is easy to check by using (ii) of lemma \ref{lem:sym} that the only terms in this sum which are nonzero are for $\xi ' = -\xi$ and $\eta'= -\eta$. One gets hence
\begin{equation}
 \left<  (\bp_0^i \bp_{\be_1}^i \bp_0^j \bp_{\be_1}^j)  \right>_{\scriptscriptstyle{N,\mathcal E}} =\cfrac{1}{N^{2d}}\sum_{\xi,\eta \neq 0} \left<\left|{\tilde \bp}^i (\xi)\right|^2 \left|{\tilde \bp}^j (\eta)\right|^2 \right>_{\scriptscriptstyle{N,\mathcal E}} \exp\left(\cfrac{2i\pi \be_1}{N} \cdot (\xi + \eta)\right)
\end{equation}
Using classical equivalence of ensembles estimates (\cite{df}), one obtains
\begin{equation}
\left<  (\bp_0^i \bp_{\be_1}^i \bp_0^j \bp_{\be_1}^j)
\right>_{\scriptscriptstyle{N,\mathcal E}} =\cfrac{\mathcal E^2}{d^2} \left(
  \cfrac{1}{N^d} \sum_{\xi \neq 0} e^{\cfrac{2i \pi \be_1}{N}\cdot
    \xi}\right)^2 + O(N^{-d}) = O(N^{-d}) 
\end{equation}

iii) By using the symmetry properties, we have
$$
\left<{\tilde q}^j(\xi) {\tilde q}^j (\xi') {\tilde \bp}^j (\eta)
  {\tilde \bp}^j (\eta') \right>_{\scriptscriptstyle{N,\mathcal E}} =0
$$
for $\xi \neq -\xi'$ or $\eta \neq -\eta'$. Hence one has
\begin{equation*}
  \begin{split}
    \left<q_{\bx}^j q_\bz^j (p_{0}^j)^2 \right>_{\scriptscriptstyle{N,\mathcal E}} \\
= \frac 1{N^{3d}}
    \sum_{\xi,\xi',\eta,\eta' \neq 0} \left<{\tilde q}^j (\xi) {\tilde
        q}^j (\xi') {\tilde \bp}^j (\eta) {\tilde \bp}^{j}(\eta')
    \right>_{\scriptscriptstyle{N,\mathcal E}} \frac{\exp{(-2i\pi
        (\xi\cdot \bz +\xi' \cdot \by )/N)}}{\omega(\xi)
      \omega(\xi')}\\ 
     = \frac 1{N^{3d}} \sum_{\xi,\eta \neq 0} \left<\left|{\tilde q}^j
         (\xi) {\tilde \bp}^i (\eta)\right|^2
     \right>_{\scriptscriptstyle{N,\mathcal E}} \frac{\exp{(-2i\pi
         \xi\cdot (\bz-\by) /N)}}{\omega(\xi)^2}\\ 
 = \frac {1}{N^{2d}}
    \sum_{\xi \neq 0} \left<\left|{\tilde q}^j(\xi)
      {\tilde \bp}^j (\be_1)\right|^2
  \right>_{\scriptscriptstyle{N,\mathcal E}} \frac{\exp{(-2i\pi
      \xi\cdot (\bz -\by) /N)}}{\omega(\xi)^2} 
  \end{split}
\end{equation*}
Estimates of  \cite{df} give
\begin{equation*}
  \left| \left<({\tilde q}^j(\xi))^2
    ({\tilde \bp}^j (\be_1))^2 \right>_{\scriptscriptstyle{N,\mathcal E}} - %\left<(\tilde q(\xi)^i)^2\right>_{\scriptscriptstyle{N,\mathcal E}} 
  \left(\frac {\mathcal E}{d}\right)^2 \right| \le \frac C{N^d}
\end{equation*}
It follows that
\begin{equation*}
\left<\bq_{\by}^j (\bq_{-\be_1}^j -\bq_{\be_1}^j) (\bp_{0}^j)^2
\right>_{\scriptscriptstyle{N,\mathcal E}}  = \frac {\mathcal E^2}{d N^{2d}}
    \sum_{\xi \neq 0} \frac{e^{-2i\pi \xi\cdot (-\be_1 -\by)
        /N}-e^{-2i \pi \xi \cdot (\be_1 -\by)/N}}{\omega(\xi)^2} +R_N 
\end{equation*}
where
\begin{equation*}
|R_N| \leq C N^{-2d} \sum_{\xi \neq 0} \cfrac{|\sin(2\pi \xi^1 /N)|}{4
  \alpha \sum_{k=1}^d \sin^2 (\pi \xi^k /N)} 
\end{equation*}
To obtain iii) observe that 
\begin{equation*}
\frac {1}{ N^{2d}} \sum_{\xi \neq 0} \frac{e^{-2i\pi \xi\cdot (-\be_1
    -\by) /N}-e^{-2i \pi \xi \cdot (\be_1 -\by)/N}}{\omega(\xi)^2}=
{\Gamma}_N (y) 
\end{equation*}
and
\begin{equation*}
N^{-2d} \sum_{\xi \neq 0} \cfrac{|\sin(2\pi \xi^1 /N)|}{4 \alpha
  \sum_{k=1}^d \sin^2 (\pi \xi^k /N)} \sim 
\begin{cases}
\log N /N, \quad d=1\\
1/N, \quad d=2\\
1/N^d,\quad d\geq 3
\end{cases}
\end{equation*}
\end{proof}

%%%%%%%%%%%%%%%%%%%%%%%%%%%%%%%%
%%%%%%%%%%%%%%%%%%%%%%%%%%%%%%%%
%%%   GREEN-KUBO
%%%%%%%%%%%%%%%%%%%%%%%%%%%%%%%%%
%%%%%%%%%%%%%%%%%%%%%%%%%%%%%%%%%

\end{document}